\documentclass[3p,times,numafflabel]{elsarticle}

\usepackage{mathtools}
\usepackage{amsmath}
\usepackage{subcaption}
\usepackage{xcolor}
\usepackage{amssymb}
\usepackage[figuresright]{rotating}

\begin{document}
\begin{frontmatter}


%\title{NxM\tnoteref{label1}}
%\title{}
%\author{Name\corref{cor1}\fnref{label2}}
%\author{Y. Liu\fnref{label2}, S. Oser\fnref{label2}, A. Pradeep\fnref{label2}}
% \author{SuperCDMS collaboration}
%\ead{email address}
%\fntext[label2]{aaaaaa}

%% e.g. \dochead{17th International Conference on Dynamical Processes in Excited States of Solids}

\title{Multi-channel, multi-template event reconstruction for SuperCDMS data using machine learning}

%\include{authorlist}
%\if{0}

\author[1,2]{M.F.~Albakry} 
\author[3]{I.~Alkhatib}
\author[4,5]{D.~Alonso-González} 
%\author[6]{D.W.P.~Amaral} 
\author[7]{J.~Anczarski} 
\author[7]{T.~Aralis} %\address{SLAC National Accelerator Laboratory/Kavli Institute for Particle Astrophysics and Cosmology, Menlo Park, CA 94025, USA}
\author[8]{T.~Aramaki} 
\author[9]{I.~Ataee~Langroudy}
\author[10]{C.~Bathurst} 
\author[9]{R.~Bhattacharyya} %\address{Department of Physics and Astronomy, and the Mitchell Institute for Fundamental Physics and Astronomy, Texas A\&M University, College Station, TX 77843, USA}
\author[9]{A.J.~Biffl} %\address{Department of Physics and Astronomy, and the Mitchell Institute for Fundamental Physics and Astronomy, Texas A\&M University, College Station, TX 77843, USA}
\author[7]{P.L.~Brink} %\address{SLAC National Accelerator Laboratory/Kavli Institute for Particle Astrophysics and Cosmology, Menlo Park, CA 94025, USA}
\author[3]{M.~Buchanan} 
\author[11]{R.~Bunker}
\author[7,29]{B.~Cabrera} %\address{SLAC National Accelerator Laboratory/Kavli Institute for Particle Astrophysics and Cosmology, Menlo Park, CA 94025, USA}
\author[12]{R.~Calkins} 
\author[7]{R.A.~Cameron} %\address{SLAC National Accelerator Laboratory/Kavli Institute for Particle Astrophysics and Cosmology, Menlo Park, CA 94025, USA}
\author[7]{C.~Cartaro} %\address{SLAC National Accelerator Laboratory/Kavli Institute for Particle Astrophysics and Cosmology, Menlo Park, CA 94025, USA}
\author[4,5]{D.G.~Cerde\~no} %\address{Departamento de F\'{\i}sica Te\'orica, Universidad Aut\'onoma de Madrid, 28049 Madrid, Spain}\address{Instituto de F\'{\i}sica Te\'orica UAM-CSIC, Campus de Cantoblanco, 28049 Madrid, Spain}
\author[13]{Y.-Y.~Chang} 
\author[14]{M.~Chaudhuri} 
\author[9]{J.-H.~Chen} %\address{Department of Physics and Astronomy, and the Mitchell Institute for Fundamental Physics and Astronomy, Texas A\&M University, College Station, TX 77843, USA}
\author[15]{R.~Chen} 
\author[16]{N.~Chott} 
\author[23,12]{J.~Cooley} %\address{SNOLAB, Creighton Mine \#9, 1039 Regional Road 24, Sudbury, ON P3Y 1N2, Canada}%\address{Department of Physics, Southern Methodist University, Dallas, TX 75275, USA}
\author[10]{H.~Coombes} %\address{Department of Physics, University of Florida, Gainesville, FL 32611, USA}
\author[17]{P.~Cushman} 
\author[3]{R.~Cyna} %\address{Department of Physics, University of Toronto, Toronto, ON M5S 1A7, Canada}
\author[14]{S.~Das} %\address{National Institute of Science Education and Research, An OCC of Homi Bhabha National Institute, Jatni 752050, India}
\author[1]{S.~Dharani} %\address{Department of Physics \& Astronomy, University of British Columbia, Vancouver, BC V6T 1Z1, Canada}
\author[11]{M.L.~di~Vacri} %\address{Pacific Northwest National Laboratory, Richland, WA 99352, USA}
\author[3]{M.D.~Diamond} %\address{Department of Physics, University of Toronto, Toronto, ON M5S 1A7, Canada}
\author[10]{M.~Elwan} %\address{Department of Physics, University of Florida, Gainesville, FL 32611, USA}
\author[17]{S.~Fallows} %\address{School of Physics \& Astronomy, University of Minnesota, Minneapolis, MN 55455, USA}
\author[18,2]{E.~Fascione} %\address{TRIUMF, Vancouver, BC V6T 2A3, Canada}
\author[15]{E.~Figueroa-Feliciano} %\address{Department of Physics \& Astronomy, Northwestern University, Evanston, IL 60208-3112, USA}
\author[19]{S.L.~Franzen} 
\author[25]{A.~Gevorgian}
\author[20]{M.~Ghaith} 
\author[3]{G.~Godden} %\address{Department of Physics, University of Toronto, Toronto, ON M5S 1A7, Canada}
\author[31]{J.~Golatkar} 
\author[22]{S.R.~Golwala} 
\author[15]{R.~Gualtieri} %\address{Department of Physics \& Astronomy, Northwestern University, Evanston, IL 60208-3112, USA}
\author[23,24]{J.~Hall} 
\author[3]{S.A.S.~Harms} %\address{Department of Physics, University of Toronto, Toronto, ON M5S 1A7, Canada}
\author[15]{C.~Hays} %\address{Department of Physics \& Astronomy, Northwestern University, Evanston, IL 60208-3112, USA}
\author[25]{B.A.~Hines} 
\author[3]{Z.~Hong} %\address{Department of Physics, University of Toronto, Toronto, ON M5S 1A7, Canada}
\author[27]{L.~Hsu} 
\author[25,26]{M.E.~Huber} %\address{Department of Physics, University of Colorado Denver, Denver, CO 80217, USA}\address{Department of Electrical Engineering, University of Colorado Denver, Denver, CO 80217, USA}
\author[3]{V.~Iyer} %\address{Department of Physics, University of Toronto, Toronto, ON M5S 1A7, Canada}
\author[14]{V.K.S.~Kashyap} %\address{National Institute of Science Education and Research, An OCC of Homi Bhabha National Institute, Jatni 752050, India}
\author[3]{S.T.D.~Keller} %\address{Department of Physics, University of Toronto, Toronto, ON M5S 1A7, Canada}
\author[9]{M.H.~Kelsey} %\address{Department of Physics and Astronomy, and the Mitchell Institute for Fundamental Physics and Astronomy, Texas A\&M University, College Station, TX 77843, USA}
\author[15]{K.T.~Kennard} %\address{Department of Physics \& Astronomy, Northwestern University, Evanston, IL 60208-3112, USA}
\author[25]{Z.~Kromer} %\address{Department of Physics, University of Colorado Denver, Denver, CO 80217, USA}
\author[23]{A.~Kubik} %\address{SNOLAB, Creighton Mine \#9, 1039 Regional Road 24, Sudbury, ON P3Y 1N2, Canada}
\author[7]{N.A.~Kurinsky} %\address{SLAC National Accelerator Laboratory/Kavli Institute for Particle Astrophysics and Cosmology, Menlo Park, CA 94025, USA}
\author[9]{M.~Lee} %\address{Department of Physics and Astronomy, and the Mitchell Institute for Fundamental Physics and Astronomy, Texas A\&M University, College Station, TX 77843, USA}
\author[8]{J.~Leyva} %\address{Department of Physics, Northeastern University, 360 Huntington Avenue, Boston, MA 02115, USA}
\author[31]{B.~Lichtenberg} 
\author[12]{J.~Liu} %\address{Department of Physics, Southern Methodist University, Dallas, TX 75275, USA}
\author[17]{Y.~Liu\corref{cor1}} %\address{School of Physics \& Astronomy, University of Minnesota, Minneapolis, MN 55455, USA}
\cortext[cor1]{yanliusp@umn.edu}
\author[4,5]{E.~Lopez~Asamar} %\address{Departamento de F\'{\i}sica Te\'orica, Universidad Aut\'onoma de Madrid, 28049 Madrid, Spain\address{Instituto de F\'{\i}sica Te\'orica UAM-CSIC, Campus de Cantoblanco, 28049 Madrid, Spain}
\author[27]{P.~Lukens} %\address{Fermi National Accelerator Laboratory, Batavia, IL 60510, USA}
\author[4,5]{R.~López~Noé} %\address{Departamento de F\'{\i}sica Te\'orica, Universidad Aut\'onoma de Madrid, 28049 Madrid, Spain
\author[7]{D.B.~MacFarlane} %\address{SLAC National Accelerator Laboratory/Kavli Institute for Particle Astrophysics and Cosmology, Menlo Park, CA 94025, USA}
\author[9]{R.~Mahapatra} %\address{Department of Physics and Astronomy, and the Mitchell Institute for Fundamental Physics and Astronomy, Texas A\&M University, College Station, TX 77843, USA}
\author[19]{J.S.~Mammo} %\address{Department of Physics, University of South Dakota, Vermillion, SD 57069, USA}
\author[2]{A.J.~Mayer} %\address{TRIUMF, Vancouver, BC V6T 2A3, Canada}
\author[3]{P.C.~McNamara}
\author[28]{\'E.~Michaud} 
\author[21]{E.~Michielin} %\address{Institute for Astroparticle Physics (IAP), Karlsruhe Institute of Technology (KIT), 76344 Eggenstein-Leopoldshafen, Germany}
\author[25]{K.~Mickelson} %\address{Department of Physics, University of Colorado Denver, Denver, CO 80217, USA}
\author[9]{N.~Mirabolfathi} %\address{Department of Physics and Astronomy, and the Mitchell Institute for Fundamental Physics and Astronomy, Texas A\&M University, College Station, TX 77843, USA}
\author[9]{M.~Mirzakhani} %\address{Department of Physics and Astronomy, and the Mitchell Institute for Fundamental Physics and Astronomy, Texas A\&M University, College Station, TX 77843, USA}
\author[14]{B.~Mohanty} %\address{National Institute of Science Education and Research, An OCC of Homi Bhabha National Institute, Jatni 752050, India}
\author[14]{D.~Mondal} %\address{National Institute of Science Education and Research, An OCC of Homi Bhabha National Institute, Jatni 752050, India}
\author[9]{D.~Monteiro} %\address{Department of Physics and Astronomy, and the Mitchell Institute for Fundamental Physics and Astronomy, Texas A\&M University, College Station, TX 77843, USA}
\author[17]{J.~Nelson} %\address{School of Physics \& Astronomy, University of Minnesota, Minneapolis, MN 55455, USA}
\author[17]{H.~Neog} %\address{School of Physics \& Astronomy, University of Minnesota, Minneapolis, MN 55455, USA}
\author[11]{J.L.~Orrell} %\address{Pacific Northwest National Laboratory, Richland, WA 99352, USA}
\author[9]{M.D.~Osborne} %\address{Department of Physics and Astronomy, and the Mitchell Institute for Fundamental Physics and Astronomy, Texas A\&M University, College Station, TX 77843, USA}
\author[1,2]{S.M.~Oser} %\address{Department of Physics \& Astronomy, University of British Columbia, Vancouver, BC V6T 1Z1, Canada}\address{TRIUMF, Vancouver, BC V6T 2A3, Canada}
\author[19]{L.~Pandey} %\address{Department of Physics, University of South Dakota, Vermillion, SD 57069, USA}
\author[17]{S.~Pandey} %\address{School of Physics \& Astronomy, University of Minnesota, Minneapolis, MN 55455, USA}
\author[7]{R.~Partridge} %\address{SLAC National Accelerator Laboratory/Kavli Institute for Particle Astrophysics and Cosmology, Menlo Park, CA 94025, USA}
\author[15]{P.K.~Patel} %\address{Department of Physics \& Astronomy, Northwestern University, Evanston, IL 60208-3112, USA}
\author[28]{D.S.~Pedreros} %\address{D\'epartement de Physique, Universit\'e de Montr\'eal, Montr\'eal, Québec H3C 3J7, Canada}
\author[3]{W.~Peng} %\address{Department of Physics, University of Toronto, Toronto, ON M5S 1A7, Canada}
\author[3]{W.L.~Perry} %\address{Department of Physics, University of Toronto, Toronto, ON M5S 1A7, Canada}
\author[19]{R.~Podviianiuk} %\address{Department of Physics, University of South Dakota, Vermillion, SD 57069, USA}
\author[12]{M.~Potts} %\address{Pacific Northwest National Laboratory, Richland, WA 99352, USA}
\author[16]{S.S.~Poudel} %\address{Department of Physics, South Dakota School of Mines and Technology, Rapid City, SD 57701, USA}
\author[7]{A.~Pradeep\corref{cor2}} %\address{SLAC National Accelerator Laboratory/Kavli Institute for Particle Astrophysics and Cosmology, Menlo Park, CA 94025, USA}
\cortext[cor2]{aditi.pradeep@slac.stanford.edu}
%\ead{aditi.pradeep@slac.stanford.edu}
\author[13]{M.~Pyle} %\address{Department of Physics, University of California, Berkeley, CA 94720, USA}\address{Lawrence Berkeley National Laboratory, Berkeley, CA 94720, USA}
\author[2]{W.~Rau} %\address{TRIUMF, Vancouver, BC V6T 2A3, Canada}
%\author[6]{E.~Reid} 
\author[3]{T.~Reynolds} %\address{Department of Physics, University of Toronto, Toronto, ON M5S 1A7, Canada}
\author[4,5]{M.~Rios} %\address{Departamento de F\'{\i}sica Te\'orica, Universidad Aut\'onoma de Madrid, 28049 Madrid, Spain\address{Instituto de F\'{\i}sica Te\'orica UAM-CSIC, Campus de Cantoblanco, 28049 Madrid, Spain}
\author[25]{A.~Roberts} %\address{Department of Physics, University of Colorado Denver, Denver, CO 80217, USA}
\author[28]{A.E.~Robinson} %\address{D\'epartement de Physique, Universit\'e de Montr\'eal, Montr\'eal, Québec H3C 3J7, Canada}
\author[10]{L.~Rosado} %\address{Department of Physics, University of Florida, Gainesville, FL 32611, USA}
\author[7]{J.L.~Ryan} %\address{SLAC National Accelerator Laboratory/Kavli Institute for Particle Astrophysics and Cosmology, Menlo Park, CA 94025, USA}
\author[10]{T.~Saab} %\address{Department of Physics, University of Florida, Gainesville, FL 32611, USA}
\author[10]{D.~Sadek} %\address{Department of Physics, University of Florida, Gainesville, FL 32611, USA}
\author[13]{B.~Sadoulet} %\address{Department of Physics, University of California, Berkeley, CA 94720, USA}\address{Lawrence Berkeley National Laboratory, Berkeley, CA 94720, USA}
\author[9]{S.P.~Sahoo} %\address{Department of Physics and Astronomy, and the Mitchell Institute for Fundamental Physics and Astronomy, Texas A\&M University, College Station, TX 77843, USA}
\author[12]{I.~Saikia} %\address{Department of Physics, Southern Methodist University, Dallas, TX 75275, USA}
\author[1,2]{S.~Salehi}
\author[19]{J.~Sander} %\address{Department of Physics, University of South Dakota, Vermillion, SD 57069, USA}
\author[22]{B.~Sandoval} %\address{Division of Physics, Mathematics, \& Astronomy, California Institute of Technology, Pasadena, CA 91125, USA}
\author[3]{A.~Sattari} %\address{Department of Physics, University of Toronto, Toronto, ON M5S 1A7, Canada}
\author[16]{R.W.~Schnee} %\address{Department of Physics, South Dakota School of Mines and Technology, Rapid City, SD 57701, USA}
\author[13]{B.~Serfass} %\address{Department of Physics, University of California, Berkeley, CA 94720, USA}
\author[25]{A.E.~Sharbaugh} %\address{Department of Physics, University of Colorado Denver, Denver, CO 80217, USA}
\author[22]{R.S.~Shenoy}
\author[7]{A.~Simchony} %\address{SLAC National Accelerator Laboratory/Kavli Institute for Particle Astrophysics and Cosmology, Menlo Park, CA 94025, USA}
\author[3]{P.~Sinervo} %\address{Department of Physics, University of Toronto, Toronto, ON M5S 1A7, Canada}
\author[7]{Z.J.~Smith} %\address{SLAC National Accelerator Laboratory/Kavli Institute for Particle Astrophysics and Cosmology, Menlo Park, CA 94025, USA}
\author[18]{R.~Soni} %\address{Department of Physics, Queen's University, Kingston, ON K7L 3N6, Canada}\address{TRIUMF, Vancouver, BC V6T 2A3, Canada}
\author[7]{K.~Stifter} %\address{SLAC National Accelerator Laboratory/Kavli Institute for Particle Astrophysics and Cosmology, Menlo Park, CA 94025, USA}
\author[16]{J.~Street} %\address{Department of Physics, South Dakota School of Mines and Technology, Rapid City, SD 57701, USA}
\author[23]{M.~Stukel} %\address{SNOLAB, Creighton Mine \#9, 1039 Regional Road 24, Sudbury, ON P3Y 1N2, Canada}
\author[10]{H.~Sun} %\address{Department of Physics, University of Florida, Gainesville, FL 32611, USA}
\author[17]{E.~Tanner} %\address{School of Physics \& Astronomy, University of Minnesota, Minneapolis, MN 55455, USA}
\author[9]{N.~Tenpas}
\author[9]{D.~Toback} %\address{Department of Physics and Astronomy, and the Mitchell Institute for Fundamental Physics and Astronomy, Texas A\&M University, College Station, TX 77843, USA}
\author[25]{A.N.~Villano} %\address{Department of Physics, University of Colorado Denver, Denver, CO 80217, USA}
\author[31]{J.~Viol} 
\author[21,31]{B.~von~Krosigk} %\address{Institute for Astroparticle Physics (IAP), Karlsruhe Institute of Technology (KIT), 76344 Eggenstein-Leopoldshafen, Germany}
\author[22]{O.~Wen} %\address{Division of Physics, Mathematics, \& Astronomy, California Institute of Technology, Pasadena, CA 91125, USA}
\author[17]{Z.~Williams} %\address{School of Physics \& Astronomy, University of Minnesota, Minneapolis, MN 55455, USA}
\author[1]{M.J.~Wilson} %\address{Department of Physics \& Astronomy, University of British Columbia, Vancouver, BC V6T 1Z1, Canada}
\author[9]{J.~Winchell} %\address{Department of Physics and Astronomy, and the Mitchell Institute for Fundamental Physics and Astronomy, Texas A\&M University, College Station, TX 77843, USA}
\author[29]{S.~Yellin} 
\author[30]{B.A.~Young} 
\author[24,23,3]{B.~Zatschler} %\address{SNOLAB, Creighton Mine \#9, 1039 Regional Road 24, Sudbury, ON P3Y 1N2, Canada}%\address{Department of Physics, University of Toronto, Toronto, ON M5S 1A7, Canada}
\author[24,23,3]{S.~Zatschler} %\address{SNOLAB, Creighton Mine \#9, 1039 Regional Road 24, Sudbury, ON P3Y 1N2, Canada}%\address{Department of Physics, University of Toronto, Toronto, ON M5S 1A7, Canada}
\author[21]{A.~Zaytsev} %\address{Institute for Astroparticle Physics (IAP), Karlsruhe Institute of Technology (KIT), 76344 Eggenstein-Leopoldshafen, Germany}
\author[3]{E.~Zhang} %\address{Department of Physics, University of Toronto, Toronto, ON M5S 1A7, Canada}
\author[9]{L.~Zheng} %\address{Department of Physics and Astronomy, and the Mitchell Institute for Fundamental Physics and Astronomy, Texas A\&M University, College Station, TX 77843, USA}
\author[3]{A.~Zuniga} %\address{Department of Physics, University of Toronto, Toronto, ON M5S 1A7, Canada}
\author[3]{M.J.~Zurowski} %address{Department of Physics, University of Toronto, Toronto, ON M5S 1A7, Canada}

%\if{0}
\address[1]{Department of Physics \& Astronomy, University of British Columbia, Vancouver, BC V6T 1Z1, Canada}
\address[2]{TRIUMF, Vancouver, BC V6T 2A3, Canada}
\address[3]{Department of Physics, University of Toronto, Toronto, ON M5S 1A7, Canada}
\address[4]{Departamento de F\'{\i}sica Te\'orica, Universidad Aut\'onoma de Madrid, 28049 Madrid, Spain}
\address[5]{Instituto de F\'{\i}sica Te\'orica UAM-CSIC, Campus de Cantoblanco, 28049 Madrid, Spain}
%\address[6]{Department of Physics, Durham University, Durham DH1 3LE, UK}
\address[7]{SLAC National Accelerator Laboratory/Kavli Institute for Particle Astrophysics and Cosmology, Menlo Park, CA 94025, USA}
\address[8]{Department of Physics, Northeastern University, 360 Huntington Avenue, Boston, MA 02115, USA}
\address[9]{Department of Physics and Astronomy, and the Mitchell Institute for Fundamental Physics and Astronomy, Texas A\&M University, College Station, TX 77843, USA}
\address[10]{Department of Physics, University of Florida, Gainesville, FL 32611, USA}
\address[11]{Pacific Northwest National Laboratory, Richland, WA 99352, USA}
\address[12]{Department of Physics, Southern Methodist University, Dallas, TX 75275, USA}
\address[13]{Department of Physics, University of California, Berkeley, CA 94720, USA}
\address[14]{National Institute of Science Education and Research, An OCC of Homi Bhabha National Institute, Jatni 752050, India}
\address[15]{Department of Physics \& Astronomy, Northwestern University, Evanston, IL 60208-3112, USA}
\address[16]{Department of Physics, South Dakota School of Mines and Technology, Rapid City, SD 57701, USA}
\address[17]{School of Physics \& Astronomy, University of Minnesota, Minneapolis, MN 55455, USA}
\address[18]{Department of Physics, Queen's University, Kingston, ON K7L 3N6, Canada}
\address[19]{Department of Physics, University of South Dakota, Vermillion, SD 57069, USA}
\address[20]{College of Natural and Health Sciences, Zayed University, Dubai, 19282, United Arab Emirates}
\address[21]{Institute for Astroparticle Physics (IAP), Karlsruhe Institute of Technology (KIT), 76344 Eggenstein-Leopoldshafen, Germany}
\address[22]{Division of Physics, Mathematics, \& Astronomy, California Institute of Technology, Pasadena, CA 91125, USA}
\address[23]{SNOLAB, Creighton Mine \#9, 1039 Regional Road 24, Sudbury, ON P3Y 1N2, Canada}
\address[24]{Laurentian University, Department of Physics, 935 Ramsey Lake Road, Sudbury, Ontario P3E 2C6, Canada}
\address[25]{Department of Physics, University of Colorado Denver, Denver, CO 80217, USA}
\address[26]{Department of Electrical Engineering, University of Colorado Denver, Denver, CO 80217, USA}
\address[27]{Fermi National Accelerator Laboratory, Batavia, IL 60510, USA}
\address[28]{D\'epartement de Physique, Universit\'e de Montr\'eal, Montr\'eal, Québec H3C 3J7, Canada}
\address[29]{Department of Physics, Stanford University, Stanford, CA 94305, USA}
\address[30]{Department of Physics, Santa Clara University, Santa Clara, CA 95053, USA}
\address[31]{Kirchhoff-Institut f{\"u}r Physik, Universit{\"a}t Heidelberg, 69117 Heidelberg, Germany} 
%\fi

\begin{abstract}
SuperCDMS SNOLAB uses kilogram-scale germanium and silicon detectors to search for dark matter.  Each detector has Transition Edge Sensors (TESs) patterned on the top and bottom faces of a large crystal substrate, with the TESs electrically grouped into six phonon readout channels per face. Noise correlations are expected among a detector's readout channels, in part because the channels and their readout electronics are located in close proximity to one another.  Moreover, owing to the large size of the detectors, energy deposits can produce vastly different phonon propagation patterns depending on their location in the substrate, resulting in a strong position dependence in the readout-channel pulse shapes. Both of these effects can degrade the energy resolution and consequently diminish the dark matter search sensitivity of the experiment if not accounted for properly.  We present a new algorithm for pulse reconstruction, mathematically formulated to take into account correlated noise and pulse shape variations. This new algorithm fits $N$ readout channels with a superposition of $M$ pulse templates simultaneously --- hence termed the N$\times$M filter.  We describe a method to derive the pulse templates using principal component analysis (PCA) and to extract energy and position information using a gradient boosted decision tree (GBDT).  We show that these new N$\times$M and GBDT analysis tools can reduce the impact from correlated noise sources while improving the reconstructed energy resolution 
for simulated mono-energetic events by more than a factor of three and
for the $^{71}$Ge K-shell electron-capture peak recoils 
measured in a previous version of SuperCDMS called CDMSlite
to $<$ 50 eV from the previously published value of $\sim$100 eV.  These results lay the groundwork for position reconstruction in SuperCDMS with the N$\times$M outputs.
\end{abstract}

%\begin{keyword}



%\end{keyword}

\end{frontmatter}
% \linenumbers

\section{Introduction}
\label{Introduction}

%\subsection{Setup of problem: Energy and position -> time series}
Experimental data often take the form of time series.  For example, the SuperCDMS experiment \cite{Rau_2012,  PhysRevLett.112.241302} 
digitizes phonon signals in Transition Edge Sensors (TESs) that are deposited on the surfaces of cryogenic semiconductor detectors~\cite{AKERIB2008476}.  Some SuperCDMS detectors also digitize charge signals produced by ionization inside the detectors.  These digitized time series contain pulses whose amplitudes are proportional to the amount of energy deposited in the detector.  The pulse shapes are determined by the location of the energy deposit inside the detector, the physics of phonon propagation, TES behaviour, and the response of the electronics. 
%Similarly, the data for gravitational wave detectors also consist of time series, in this case, proportional to the strain (fractional stretching of spacetime) produced by an incident gravitational wave.  Both the amplitude and shape of the signal depend on parameters of astrophysical interest, such as the masses of coalescing black holes, their distance from the detector, and the orientation of the orbit relative to Earth.  
%
%In either case, the analyst seeks 
We seek to fit templates/shapes to the time series data to determine underlying parameters of the initiating event.  Traditionally, the SuperCDMS experiment estimates the energy deposited by particles scattering in its detectors by fitting a fixed pulse shape to the digitized time series and then converting the fitted amplitude to an equivalent energy -- the so-called optimal filter~\cite{golwalathesis}.  In this paper, we generalize this approach to the case of pulse shapes that vary with the position of the energy deposits which are measured in multiple sensors containing both independent and common-mode (correlated) noise components.  The general approach can be adapted to any experiment that analyzes time series data with varying pulse shapes and/or multiple sensors.

Section~\ref{Introduction} of the paper introduces optimal filtering and describes the complications that arise when dealing with multiple channels, pulses with {\em a priori} unknown start times, and shape variations.  Section~\ref{NxM filter} describes a computationally tractable method for addressing these complications.  The remaining sections of the paper describe how this method is applied in detail to SuperCDMS, including validation with simulation and data.

\subsection{Fitting a time series with an optimal filter}
\label{sec:simple_of}

Let $d_j$ denote a discrete time series with $S$ equally spaced time samples at times $t_j = j\Delta t$, where $\Delta t$ is the sampling interval.  Let $T(t)$ denote a template: a pulse shape of normalized amplitude that we wish to fit to the time series to determine the amplitude of the measured pulse.  In the case of Gaussian white noise, the measurement uncertainty $\sigma_j$ on each measured point is constant and statistically independent of the uncertainty on any other point.  In this restricted circumstance, one can write a simple time-domain $\chi^2$ for the fit as
\begin{equation}
\chi^2(a) = \sum_{j=1}^{S} \left( \frac{d_j - aT_j}{\sigma}\right)^2.
\label{eq:timedomainchi2}
\end{equation}
Here $T_j \equiv T(t_j)$, $a$ is an unknown amplitude we wish to fit for, and the sum is over all $S$ time samples.  The best-fit amplitude $\hat{a}$ is the value of $a$ that minimizes this $\chi^2$. In general, different points in the time series may have correlated uncertainties, i.e. their covariance matrix $V_{jk} \equiv {\rm cov}(d_j,d_k)$ is not necessarily diagonal.  In this case, one can generalize the $\chi^2$ to account for a non-diagonal covariance matrix as follows:
\begin{equation}
\chi^2(a) = \sum_{j=1}^{S} \sum_{k=1}^{S} (d_j - aT_j)(d_k - aT_k)(V^{-1})_{jk}.
\label{eq:timedomainchi2_nondiag}
\end{equation}

If the covariance matrix $V$ between measured time samples is known and if it can be efficiently inverted, this formulation may suffice.  In practice, for a time series with $S$ samples, the covariance matrix, of size $S \times S$, may be too large to practically invert.  Instead, it is usually computationally more efficient to formulate the problem in the frequency domain. Let $\tilde{D}_f$ and $\tilde{T}_f$ denote the discrete Fourier transforms of the time series $d_j$ and $T_j$. It is common for the statistical properties of the measurement noise to be invariant as a function of time (``stationary"). If the measured noise is stationary, then the Fourier amplitudes at different frequencies will be statistically independent: ${\rm cov}(\tilde{D}_{f_1}, \tilde{D}_{f_2}) = 0$ for $f_1 \ne f_2$.  In this case a $\chi^2$ expression in the frequency domain takes a simple form:
\begin{equation}
\chi^2(a) = \sum_{f \ge 0} \frac{|\tilde{D}_f - a\tilde{T}_f|^2}{\sigma_f^2}.
\label{eq:freqdomainchi2}
\end{equation}
Here the sum is over frequencies $f = (1/\Delta t) k$, where $k = 0, 1, ... ~(S/2)$, and the term in the denominator, $\sigma_f^2$, is the power spectral density (PSD) of the measurement noise.  Note as well that although a Fourier transform yields both positive- and negative-frequency components, for a real time series $d_j$ the negative-frequency components of its Fourier transform are redundant with their positive frequency counterparts, related through complex conjugation $(\tilde{D}_{-f} = \tilde{D}_f^*)$.  Hence, the sum in the $\chi^2$ can be restricted to non-negative frequencies with no loss of information.  

Working in the frequency domain inherently requires the use of complex numbers, yet their incorporation into the $\chi^2$ to yield a real value must be done with care.  
The form of the numerator in Equation~\ref{eq:freqdomainchi2} can be better understood by recalling that, for a complex number $z$, $|z|^2 = [\Re{(z)}]^2 + [\Im{(z)}]^2$.  For $z = \tilde{D}_f - a\tilde{T}_f$ being the difference between the data and scaled template in the frequency domain, Equation~\ref{eq:freqdomainchi2} can be interpreted as saying that the real and the imaginary components of this difference can be treated as two statistically independent data points that each contribute to the $\chi^2$.  The statistical independence of the real and imaginary components follows from the fact that the noise is assumed to be stationary.  
One can also understand that nothing is lost in restricting the sum in the $\chi^2$ to positive frequencies: the number of independent real and imaginary degrees of freedom in $\tilde{D}_f$ for $f \ge 0$ is exactly equal to the number of time samples $S$ in the time domain\footnote{One might worry that because there are $S/2 + 1$ frequencies in the sum, each of which contributes a real and an imaginary part, Equation~\ref{eq:freqdomainchi2} contains $S + 2$ random variables.
    However for a real time series the first and last terms in the sum will always have imaginary components of zero, i.e. $\Im{(D_0)} = \Im{(D_{f_\mathrm{max}})} = 0$, thus reducing the number of free components to the expected $S$.}.  
    
Because Equation~\ref{eq:freqdomainchi2} is quadratic in $a$, the value of $a$ that minimizes the $\chi^2$ can be found analytically by taking its derivative with respect to $a$ and setting it equal to zero, yielding
\begin{equation}
\hat{a} = \frac{\sum_f \tilde{T}_f^* \tilde{D}_f / \sigma_f^2 }
{\sum_f \tilde{T}_f^* \tilde{T}_f / \sigma_f^2 }    .
\label{eq:1x1OFamp}
\end{equation}
This expression for $\hat{a}$ is referred to as the optimal filter (OF) estimate for the amplitude $a$.  It is optimal since, by the Gauss-Markov theorem, it is an unbiased estimator for $a$ with the least possible statistical variance of all linear functions of the data.  

\subsection{Dealing with time offsets between the data and the template}
\label{sec:time_shift}

The above discussion assumed that the fitted template $T(t)$ is aligned in time with the pulse in the time series data.  If however the template is offset by some time shift $\delta$ relative to the pulse, one can account for the shift by making the template a function of $\delta$, writing $T_j(\delta) = T(t_j - \delta)$ in the time domain, or $\tilde{T}_f(\delta) = {\rm DFT}[T_j(\delta)]$ for the frequency domain version, where DFT is the Discrete Fourier Transform. This shifted template $\tilde{T}_f(\delta)$ can be substituted into Equation~\ref{eq:freqdomainchi2}, with the $\chi^2$ then becoming a function of both $a$ and $\delta$.  Unfortunately, the dependence on $\delta$ is non-linear, and the minimum cannot be found analytically.  It can however be found numerically by scanning over a range of $\delta$ and minimizing the $\chi^2$ at any particular value of $\delta$ with respect to $a$.  Equation~\ref{eq:1x1OFamp} still holds provided that one substitutes $\tilde{T}_f(\delta)$ for $\tilde{T}_f$, with 
$\hat{a}$ now becoming a function of $\delta$. The best-fit $\hat{a}$ is then the one corresponding to the global minimum of $\chi^2$ with respect to $a$ and $\delta$.

\begin{figure}
    \centering
    \includegraphics[trim = 0cm 5cm 0cm 8cm, width=4in]{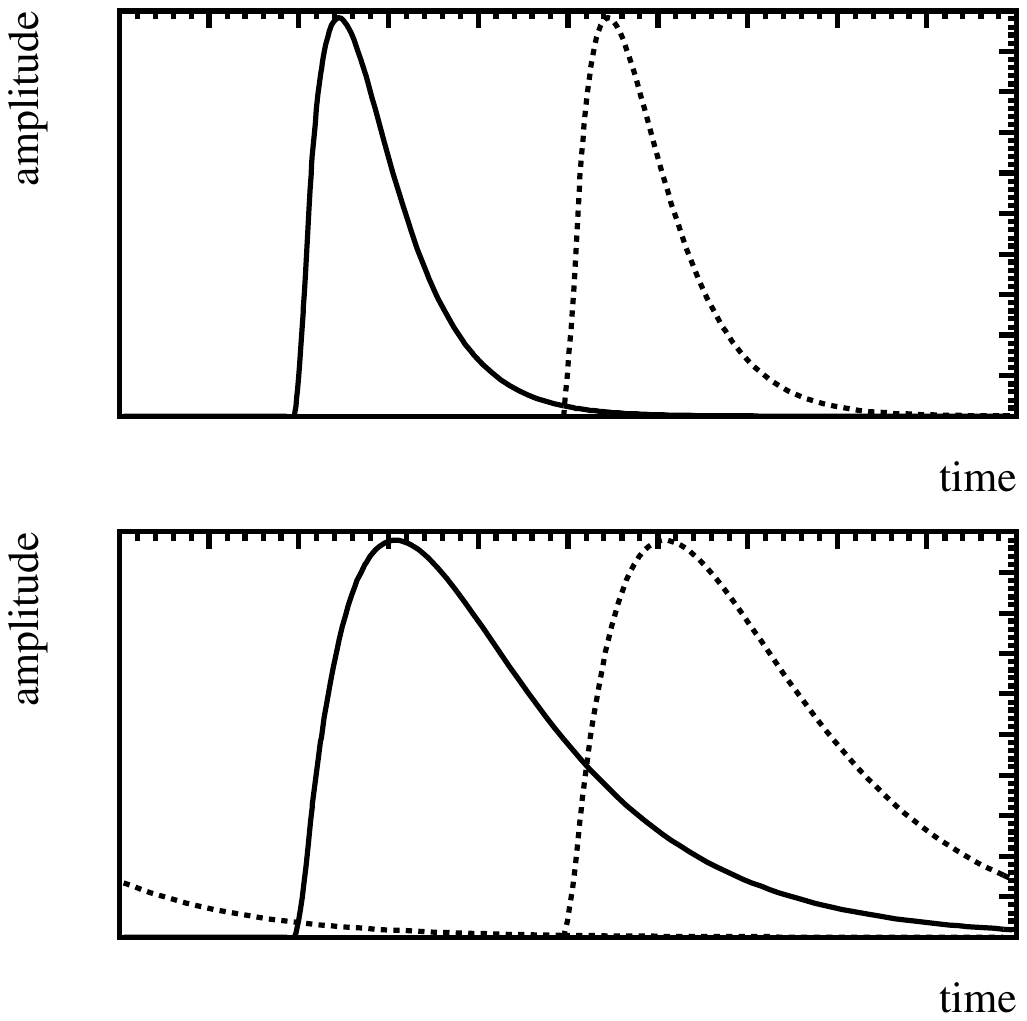}
    \caption{Top: example of a template and time shift (solid to dashed line) that can safely be handled using Equation~\ref{eq:timeshift_template}.  Bottom: a problematic template for which Equation~\ref{eq:timeshift_template} fails, because the tail of the template gets wrapped around to the front of the template (dashed line) when the template is shifted to the right.}
    \label{fig:wraparound}
\end{figure}

For the case of $\delta$ being an integer multiple of the sampling period $\Delta t$, $\tilde{T}_f$ has a particularly simple form:
\begin{equation}
\tilde{T}_f(\delta) = e^{-2\pi if \delta} \tilde T_f(0).    
\label{eq:timeshift_template}
\end{equation}
In this case, the Fourier transform of the time-shifted template can be readily obtained from the Fourier-transformed unshifted template simply by rotating each component by a complex phase $-2\pi f\delta$.  It is important to emphasize that this relation only holds for the case that $\delta$ is an integer multiple of $\Delta t$.  Moreover, Equation~\ref{eq:timeshift_template}, which can be derived from the expression of the discrete Fourier transform, implicitly assumes that $T(t)$ is periodic with period $S \Delta t$.  It has the effect of ``wrapping around'' the template --- any part of the template shifted beyond the end of the sampling period appears back at the start of the template, as shown in the bottom part of Figure~\ref{fig:wraparound}.  This does not cause problems if the templates go to zero at both ends (top of Figure~\ref{fig:wraparound}), but can result in template discontinuities otherwise (bottom of Figure~\ref{fig:wraparound}).  In cases where the time shift is not an integer multiple of $\Delta t$ or where Equation~\ref{eq:timeshift_template} introduces wraparound effects for time shifts of interest, one must instead directly calculate $T(t_j - \delta)$ and explicitly take its Fourier transform.

\subsection{Dealing with pulse shape variation}
\label{Dealing with pulse shape variation}

Section~\ref{sec:simple_of} assumed a fixed pulse shape with no variation, for which the overall amplitude was the only free parameter.  If the template shape $T(t)$ assumed in the fit is not an accurate representation of the average pulse shape, a simple one-parameter fit for the amplitude will typically be biased.  %More generally the pulse shape may vary as a function of other parameters.  
If the pulse shape varies from event to event, then fitting every event with a single fixed pulse shape will tend to broaden the amplitude resolution due to the unmodeled variation in shape.  In the context of the SuperCDMS experiment, dependence of the phonon pulse shape on the position of the energy deposit inside the detector produces a noticeable smearing of the energy resolution if no effort is made to account for the position variation. 
%\textcolor{blue}{ Shouldn't this be: In the context of the SuperCDMS experiment, dependence of the phonon pulse shape on the position of the energy deposit inside the detector produces a noticeable smearing of the energy resolution if no effort is made to account for the position variation?}  Yes---made that change.

The approach in Section~\ref{sec:simple_of} can be readily generalized to the case of template shapes that depend on additional parameters.  In this case the $\chi^2$ can be minimized with respect to all free parameters.  Provided that the parameterized templates accurately model the variation of pulse shapes in the data, the additional parameters provide additional degrees of freedom that help reduce the deviation between the fitted and true pulse shape.

If the dependence of the pulse shape model on the fitted parameters is not linear, the best-fit parameters that minimize the $\chi^2$ cannot generally be solved for analytically, although numerical minimization is usually possible.  The addition of a relative time shift between the data and fitted template, described in Section~\ref{sec:time_shift}, is an example in which the model is linear in one parameter (amplitude $a$) but non-linear in a second (time shift $\delta$). %In analysis of data from gravitational wave detectors, the ``F-statistic'' approach yields a fit which is linear in a combination of four source parameters (amplitude, inclination angle, polarization direction, and phase) and non-linear in other parameters (frequency, frequency derivation, sky position)~\cite{PhysRevD.58.063001}.
In some cases, a simple reparameterization can make a problem linear. For example, fitting a sine wave with unknown phase and amplitude can either be done as a non-linear fit $T(t) = a \sin (\omega t + \phi)$ for $a$ and $\phi$, or equivalently as a linear fit $T(t) = A \sin \omega t + B \cos \omega t$ for $A$ and $B$.

An alternative to attempting to fit for a varying pulse shape is to consider any variation in pulse shape to be a systematic uncertainty in the fit and to include that systematic uncertainty as part of the covariance matrix.  For example, the covariance matrix $V_{jk}$ for time samples  in Equation~\ref{eq:timedomainchi2_nondiag} can be expressed as a noise contribution and a contribution from pulse shape variation:
\begin{equation}
{\bf V} = {\bf V}_\mathrm{noise} + {\bf V}_\mathrm{shape}.     
\label{eq:cov_total}
\end{equation}
Here, ${\bf V}_\mathrm{shape}$ contains the covariances between time bins due to any pulse shape variation.  Equation~\ref{eq:cov_total} can be substituted into Equation~\ref{eq:timedomainchi2_nondiag} and the resulting $\chi^2$ is minimized.  This approach does not attempt to fit for any parameters that describe the shape of the pulse, but effectively ``deweights'' those features of the time series that are uncertain due to the underlying variation in pulse shapes. The additional contribution to the covariance matrix can be thought of as an additional uncertainty that will tend to broaden the fitted distribution of $a$ relative to the case in which there is no variability in pulse shape. However, the broadening from including $\bf{V}_\mathrm{shape}$ is typically smaller than the broadening in resolution caused by fitting a fixed template shape to pulses that do not all follow that shape.

While stationary noise results in a diagonal noise matrix expressed in the frequency domain, the additional covariance contribution from ${\bf V}_\mathrm{shape}$ can have any form, and in general the combined covariance matrix will no longer be diagonal when the fit is reformulated in the frequency domain.  That is, the pulse shape uncertainty can be thought of as a source of non-stationary noise.  Equation~\ref{eq:freqdomainchi2}, which assumes that there is no covariance between the measurement uncertainties of the real and imaginary parts of the Fourier amplitudes at different frequencies, thus does not hold, and there is little benefit from working in the frequency domain rather than the time domain in this case.  

The SuperCDMS experiment has traditionally used a combination of Equations~\ref{eq:timedomainchi2_nondiag} and \ref{eq:cov_total} as its most reliable simple energy estimator, and refers to this as a ``non-stationary optimal filter''~\cite{pylethesis, thakurthesis}.  This combination is found to give better energy resolution than a simple stationary optimal filter which assumes a fixed pulse shape.

Computationally, approaches in which the pulse shape variation can be captured by a linear function of a small number of fit parameters are much more efficient than either a non-linear fit or attempting to absorb the effects of shape variation as an additional uncertainty that results in a non-diagonal covariance matrix.

\subsection{Dealing with multiple channels}

Data may consist of multiple time series, either from separate detectors (e.g. jointly fitting gravitational wave data from physically separate detectors) or from multiple channels within a detector (e.g. the segmented phonon sensors on a SuperCDMS detector, each covering a different portion of the detector's surface). If the time series from different channels are aligned in time (or can be straightforwardly aligned by shifting to correct for ``time of flight'' effects), a simplistic approach may be to sum the time series and then fit the summed pulse:
\begin{equation}
d_{\mathrm{total},j} = \sum_{c=1}^{N} w_{c} d_{c,j},    
\label{eq:sum_timeseries}
\end{equation}
where $d_{c,j}$ is the time series data for channel $c$, $w_c$ is a weight factor for that channel, and $N$ is the number of channels.  This summed time series $d_{\mathrm{total},j}$ can then be used in any of the fits given in previous sections.  This approach, while easy to implement, is not guaranteed to give optimal results, especially if the weights $w_c$ are not chosen carefully.  

SuperCDMS has traditionally calculated its energy estimators using the summed time series from all of a detector's phonon channels.
%SuperCDMS has traditionally used the summed time series from all of a detector's phonon channels as the data to which it applies its energy estimators.
Not only is this approach easy, it also tends to mitigate (although not eliminate) energy resolution broadening due to position dependence across the detector, because adding the signals from all channels tends to average away this position dependence.  This is especially true in the non-stationary optimal filter, where a summed pulse is created first to mitigate position dependence to first order and then the rest of the position dependence is treated as correlated (non-stationary) noise~\cite{pylethesis}.

A better approach is to form a $\chi^2$ as in Equation~\ref{eq:freqdomainchi2} for each channel, and then to sum these $N$ $\chi^2$ terms to form a combined $\chi^2$ that can be minimized as a function of any free parameters.  Imagine that each channel has a separate template $T_c$ and amplitude $a_c(E)$ that depends on a free parameter $E$, which we can think of as being an energy.  We can then generalize Equation~\ref{eq:freqdomainchi2} to the case of multiple channels as

\begin{equation}
\chi^2(E) = \sum_{c=1}^{N} \chi_c^2 = 
\sum_{c=1}^{N} \sum_{f \ge 0} \frac{|\tilde{D}_{c,f} - a_c(E)\tilde{T}_{c,f}|^2}{\sigma_{c,f}^2}.
\label{eq:summing_chi2}
\end{equation}
This expression accounts for the possibility that the template shape, amplitude, and noise PSD $\sigma_f^2$ may differ from channel to channel.

The approach of simply adding the $\chi^2$ expressions from separate channels implicitly assumes that there are no correlated uncertainties between different channels.  
%This may be a good approximation when combining data from physically distant gravitational wave detectors.  
However, it may not be appropriate when combining data from different channels inside one detector.  For example, noise induced by mechanical vibrations or electronic interference is likely to appear as correlated noise across all channels, and in this case Equation~\ref{eq:summing_chi2} will not correctly account for these correlations.  Simply summing the raw time series as per Equation~\ref{eq:sum_timeseries} has the same issue: while uncorrelated noise will tend to average out in the sum, noise that is correlated between channels will not cancel in the sum and does not benefit from any averaging effect.

To properly account for correlated noise, one must include the covariances between measured quantities.  However, in the context of Equation~\ref{eq:freqdomainchi2}, in which the measured quantities are expressed as the complex Fourier amplitudes $\tilde{D}_f$, particular care must be taken.  Consider two complex random variables $z_1$ and $z_2$. Fully characterizing the covariances of these variables requires ten different real numbers:
\begin{equation}
\begin{bmatrix}
{\rm cov}(\Re(z_1),\Re(z_1)) & {\rm cov}(\Re(z_1),\Im(z_1)) & {\rm cov}(\Re(z_1),\Re(z_2)) & {\rm cov}(\Re(z_1),\Im(z_2))  \\    
{\rm cov}(\Im(z_1),\Re(z_1)) & {\rm cov}(\Im(z_1),\Im(z_1)) & {\rm cov}(\Im(z_1),\Re(z_2)) & {\rm cov}(\Im(z_1),\Im(z_2))  \\    
{\rm cov}(\Re(z_2),\Re(z_1)) & {\rm cov}(\Re(z_2),\Im(z_1)) & {\rm cov}(\Re(z_2),\Re(z_2)) & {\rm cov}(\Re(z_2),\Im(z_2))  \\    
{\rm cov}(\Im(z_2),\Re(z_1)) & {\rm cov}(\Im(z_2),\Im(z_1)) & {\rm cov}(\Im(z_2),\Re(z_2)) & {\rm cov}(\Im(z_2),\Im(z_2))  \\   
\end{bmatrix}. 
\label{eq:realmatrix}
\end{equation}

In contrast, problems arise if one tries to directly calculate covariances between complex numbers, i.e. ${\rm cov}(z_i,z_j)$.  The proper definition of variance of a complex random variable is ${\rm E}[|z - {\rm E}[z]|^2]$ rather than ${\rm E}[(z - {\rm E}[z])^2]$, % Scott overrules Ray's suggestion to delete the previous clause "rather than ..."
and is thus a real number~\cite{complexvariablesstatistics}.  The covariance of two random variables is defined as ${\rm E}[(z_1 - {\rm E}[z_1])(z_2 - {\rm E}[z_2])^*] = {\rm E}(z_1 z_2^*) - {\rm E}(z_1){\rm E}(z_2^*)$.  A $2\times 2$ covariance matrix $V_{ij} = {\rm cov}(z_1,z_2)$ formed using these definitions from two complex numbers can be parameterized by only four real numbers, illustrating that this covariance matrix does not fully capture all of the possible correlations.  To fully characterize all of the covariances between complex random variables, one must also include pseudo-covariances, defined as ${\rm pseudocov}(z_1,z_2) \equiv {\rm cov}(z_1,z_2^*) = {\rm E}[(z_1 - {\rm E}[z_1])(z_2 - {\rm E}[z_2])]$.  Together, the $2 \times 2$ covariance and pseudo-covariance matrices fully capture all possible covariances among components of the complex variables.  Unfortunately, the form of a likelihood or $\chi^2$ in this general case is quite complicated~\cite{complexvariablesstatistics}.

In the special case of complex variables that have circular symmetry (i.e. are uniformly distributed in their complex phase), the pseudo-covariance matrix will be identically zero, and there is a simple form for the $\chi^2$ including correlations among channels as was derived in \cite{thakurthesis}:
\begin{equation}
\chi^2(a) = \sum_f \sum_{c_1} \sum_{c_2}  (\tilde{D}_{c_1,f} - a\tilde{T}_{c_1,f}) (\tilde{D}_{c_2,f} - a\tilde{T}_{c_2,f})^* V^{-1}_{c_1 c_2,f}.
\label{eq:complexchi2}
\end{equation}
Here, the summation is over frequency bins and over channels indexed by $c_1$ and $c_2$.  Note that the covariance matrix $V_{ij,f} = {\rm cov}(\tilde{D}_{c_1,f},\tilde{D}_{c_2,f})$ is an $N \times N$ covariance matrix that is a function of frequency, and that $V_{ij,f} = V_{ji,f}^*$ (ensuring that the above expression is purely real).  Equation~\ref{eq:complexchi2} assumes that the noise is uncorrelated between frequencies, i.e. at any given frequency the noise may be correlated between channels, but the noise at any given frequency is assumed to be uncorrelated with the noise at any other frequency.

In order to avoid relying on the assumption of circularly symmetric noise, and to avoid the complexities of dealing with both covariance and pseudo-covariance matrices, in this paper we will formulate our $\chi^2$ expressions using only real numbers, treating each complex random variable as two real random variables.  Therefore, if we have a vector of $N$ complex amplitudes, we rewrite this as a real vector with $2N$ components and express their covariances with a $2N \times 2N$ matrix along the lines of Equation~\ref{eq:realmatrix}.

\subsection{The approach in this paper}

A general solution to the problems of pulse shape variation and correlated noise is to formulate a generic $\chi^2$ that encodes all possible correlations, and to allow the fitted signal templates to be functions (possibly non-linear) of the fit parameters.  Unfortunately, both aspects of this approach present considerable computational difficulties.  With $S$ samples and $N$ channels of data, the covariance matrix capturing any possible correlations between data points would have a size of $S N \times S N$.
%\textcolor{blue}{I would have preferred to write this in 3D as N$\times$N$\times$S}.  That has the wrong dimensions---with S samples and N channels, the number of possible correlations is of order (NS)^2 not (N^2)S.
For anything but very short time series the inversion of this matrix will likely be computationally intractable.%, since the number of required operations scales with the matrix size cubed.

In this paper, we make the simplifying assumption that noise is uncorrelated between frequencies but allow for correlations between channels at any given frequency.  This simplification effectively means that rather than having to invert a single matrix of size $S N \times S N$, we instead have to invert $\mathcal{O}$($S/2$) matrices of size $2N \times 2N$ (one for each non-negative frequency in the discrete Fourier transform), where each covariance matrix is a purely real matrix along the lines of Equation~\ref{eq:realmatrix}, encoding the correlations among the real and imaginary components of $\tilde{D}_f$.  Due to the much smaller matrix size, this approach is computationally tractable, even if it must be done for each frequency.

Accounting for pulse shape variation requires a different approach.  An accurate parameterization of the pulse shape as a function of the underlying physics parameters would be ideal, but developing this parameterization may be non-trivial, especially if the detector physics is complicated and not analytically calculable.  Furthermore, if the pulse shape varies non-linearly with these parameters, then the $\chi^2$ cannot generally be minimized analytically, increasing the computational load.  In this paper, we take the approach of empirically modelling the varying pulse shapes as linear combinations of different templates, derived from data, which capture the most significant variations.  The end result is that we fit time series from $N$ channels with linear combinations of $M$ different templates, resulting in a total of $NM$ fitted amplitudes\footnote{Throughout this paper we use $N\times M$ to refer to the fitter, while $NM$ is the number of fitted amplitudes.}.  Section~\ref{NxM filter} describes the mathematical formalism to perform this fit while taking into account the measured covariances between channels at each frequency.

The $N \times M$ amplitudes encode information about the characteristics of the underlying pulse, such as its energy and position.  However, the parameterization employed is empirical, and we need to relate the fitted amplitudes to parameters of physical interest, such as the energy and position of the event inside a SuperCDMS detector.  Section~\ref{Applications in SuperCDMS} describes how machine learning techniques can be used to extract meaningful pulse templates and to map the fitted amplitudes to the energy and position of the energy deposit inside the detector.

\section{The N$\times$M optimal filter}
\label{NxM filter}

%\subsection{Mathematical formulation}
%\label{Mathematical formulation}

Suppose that we have $N$ channels to which we fit $M$ templates with different shapes. We can write
the prediction $p_{c,s}$ in the time domain of the data of channel $c$ for the 
$s^\textrm{th}$ time bin sample as:
\begin{equation}
p_{c, s} = \sum_{i=1}^M a_{ci} T_{ci, s}(\delta_{ci}) = \sum_{i=1}^M a_{ci} T_{ci, s}(\delta_{c}).
\label{eq:basicmodel}
\end{equation}
Here, the sum over $i$ denotes a sum over templates $T_{ci,s}$, and
$a_{ci}$ and $\delta_{ci}$ are the amplitude and time delay for the $i^\textrm{th}$ template. 
From a physics standpoint, we expect that the templates of the same channel share the same starting point, which is dependent on the event interaction time and signal propagation speed. Consequently, in the last equality we impose the constraint that templates of the same channel have identical time delays, thus $\delta_{ci} = \delta_{c}$. 

Similar to the single-channel, single-template OF case, the $\chi^2$ has a nonlinear dependence on $\delta_{c}$. As the number of channels increases (beyond four), the $\chi^2$ minimization involving $\delta_{c}$ becomes computationally challenging. For instance, if each of twelve channels allows five possible time-delay values, a brute-force grid search would require evaluating $5^{12}$ combinations. We address this issue by first extracting the time delay for each channel using standard single-channel, multiple-template optimal filter. This process can be solved analytically using an inverse Fourier transform, as described in \cite{PhysRevD.85.122006}. Ideally, the template would be shifted by the amount of the time delay. However, we opted to shift the pulse data instead to reduce memory consumption. For time delay shifts smaller than 40~$\mu$s, the two operations are effectively equivalent for frequencies below 25~kHz, where the majority of the pulse information resides. %under the assumption that the frequencies are uncorrelated. %\textcolor{purple}{YL: I reworded above.}

Equation \ref{eq:basicmodel} implies an equal number of templates ($M$) across channels, albeit this is primarily a mathematical formality. In practice, one can first identify the maximum number of templates required for any channel and set $M$ to this maximum number. Channels with fewer templates can be accommodated by including additional templates whose amplitudes are fixed to be zero.
 
The final output of the N$\times$M optimal filter is therefore $NM$ amplitude coefficients $a_{ci}$. The correlation among these coefficients can be very complicated, and the ultimate energy or position estimators will be expressed in a non-linear mapping of $a_{ci}$, as demonstrated in Section~3. %Furthermore, external constraints can also be imposed on $a_{ci}$ prior to the fit. For example, we could require that a template has the same amplitude in all channels, or we may want to set some of the coefficients to zero.

Equation \ref{eq:basicmodel} can be written in matrix form as
\begin{equation}
\vec{p}_c = \vec{a}_{c} \cdot {\bf {T}}_c(\delta_c),
\label{eq:modelmatrixform}
\end{equation}
where $\vec{p}_{c}$ is a $1 \times S$ vector containing the predicted time series of channel $c$, $\vec{a_{c}}$ is an amplitude vector of size $1 \times M$, and ${\bf {T}}_c$ is a template matrix of the size $M$ $\times$ $S$ containing the $M$ templates for channel $c$, each containing $S$ samples. Taking a Fourier transform of Equation~\ref{eq:modelmatrixform} and expressing the time shift of the templates explicitly using Equation~\ref{eq:timeshift_template} (for a time shift adequately approximated by an integer number of sampling periods) yields
\begin{equation}
\vec{\tilde{P}}_{c} = \vec{a}_c  \cdot {\bf \tilde{T}}_{c} \cdot {\bf R}(\delta_c).
\label{eq:ptodata}
\end{equation}
Here, $\vec{\tilde{P}}_c$ is the Fourier transform of the predicted time series for channel $c$, with a length given by the number $F$ of independent frequencies in the Fourier transform\footnote{For $S$ even, $F = S/2 + 1$, as discussed in Section~\ref{sec:simple_of}.}.  The $M \times F$ matrix $\bf {\tilde{T}}_c$ contains the Fourier transforms of the $M$ templates evaluated at a time shift of $\delta_c = 0$.  $\bf R(\delta_c)$ is an $F \times F$ diagonal matrix that applies the time shift to the templates by rotating each frequency component by $e^{-2\pi if\delta_c}$, i.e. $R_{f_1 f_2} = \delta_{f_1 f_2} e^{-2\pi i f_1 \delta_c}$.  All of the matrices in Equation~\ref{eq:ptodata} are complex matrices.

%$\hat{\vec{\tilde{D_{c}}}}$ and ${\bf \tilde{T}}_{c}$ are complex vector and matrix. ${\bf R}(t_c)$ is a rotation matrix that rotates the template ${\bf \tilde{T}}_{c}$ in the complex plane by a phase $\phi = 2\pi f t_c$. This is
%equivalent to doing a mapping such as:
%\begin{displaymath}
%{\rm Re}(\tilde{T}_{ci}) \to \cos \phi~ {\rm Re}(\tilde{T}_{ci}) +\sin \phi~ {\rm Im}(\tilde{T}_{ci})
%\end{displaymath}
%\begin{displaymath}
%{\rm Im}(\tilde{T}_{ci}) \to -\sin \phi~ {\rm Re}(\tilde{T}_{ci}) + \cos \phi~ {\rm %Im}(\tilde{T}_{ci}).
%\end{displaymath}
%for every template of channel $c$. Note that ${\bf R}(t_c)$ is independent of %template and frequency.

Let us define the difference between measurement and prediction as $\Delta$, so that at some frequency $f$
\begin{equation}
\Delta_{c, f}~=~\tilde{D}_{c, f}~-~{\tilde{P}}_{c, f}~=~\tilde{D}_{c,f}~-~(\vec{a}_{c} \cdot {\bf \tilde{T}}_{c} \cdot {\bf R}(\delta_c))_f .
\end{equation}
The $\chi^2$ for channel $c$ can be written as
\begin{equation}
\chi^2_{c} = \sum_f\chi^2_{c, f} = \sum_f \frac{|\Delta_{c, f}|^2}{\sigma^2_{c,f}} = \sum_f \frac{[\Re{(\Delta_{c,f})]}^2 + [\Im{
    }(\Delta_{c,f})]^2}{\sigma^2_{c,f}}
\end{equation}
where $\sigma^2_{c,f}$, a scalar, is the (variance) noise of channel $c$ at frequency $f$. Expanding the $\chi^2$ to include multiple channels means that the correlations between different channels, as well as between the real and imaginary components of the same channel, need to be taken into account. %These correlations can be expressed in a covariance matrix in place of $\sigma^2_{c,f}$.

For a given frequency, a real covariance matrix of
size $2N$ $\times$ $2N$ is needed to encode the covariances among the real and imaginary components of the data from the $N$ channels.  Note that some off-diagonal elements of this matrix may be zero, e.g. between the real and imaginary terms of the same channel if the noise is stationary\footnote{The general observation from data is that these terms are close to but not exactly zero -- either because the stationary noise assumption is not strictly met or because of statistical variation.}.  Let us denote this covariance matrix as ${\bf V}_{j\alpha, k\beta}$, where $\alpha$ and $\beta$ can represent either the real or imaginary parts of the data for the channels $j$ and $k$,
respectively.  Likewise, we denote the real and imaginary parts of
$\Delta_c$ with a Greek subscript.  The total $\chi^2$ can then be expressed as
\begin{equation}
\chi^2 = \sum_f \chi^2_f = \sum_f \sum_{j=1}^N \sum_{k=1}^N \sum_{\alpha = {\rm Re, Im}}
\sum_{\beta = {\rm Re, Im}} \Delta_{j\alpha,f}   \Delta_{k\beta,f} ({\bf V}^{-1})_{j\alpha,k\beta,f}.
\label{eq:chi2f}
\end{equation}
Ignoring the sum over frequencies, the remaining summations are over channels (indices $j$ and $k$) and the real/imaginary components of the Fourier amplitudes (indices $\alpha$ and $\beta$). Note that this expression corresponds to the inverse covariance matrix ${\bf V}_f^{-1}$ (dimension
$2N \times 2N$, an entirely real matrix) multiplied on either side by
a vector of length $2N$ representing the real and complex components of
the residuals $\Delta_f$.  Equation~\ref{eq:chi2f} accounts for all correlated noise between channels at any frequency, but assumes that noise at different frequencies is uncorrelated.

The exact format of $\Delta$ and ${\bf V_{\it f}}$ in Equation~\ref{eq:chi2f} can be arbitrarily defined in principle. Here we illustrate one possible option which writes everything in terms of purely real quantities. Let $\tilde{D}_f$ be a row vector of length $2N$, which contains data from all channels at frequency $f$ in sequential order, alternating between real and imaginary components. Its transpose $\tilde{D}_f^T$ looks like
\begin{equation}
\tilde{D}_f^T = \begin{pmatrix}
\Re({\rm chan1~FFT}) \\
\Im({\rm chan1~FFT}) \\
\Re({\rm chan2~FFT}) \\
\Im({\rm chan2~FFT}) \\
\vdots \\
\Re({\rm chanN~FFT}) \\
\Im({\rm chanN~FFT}) \\
\end{pmatrix}.
%\label{eq:dvecdef}
\end{equation}
The transpose of the corresponding $\tilde{P}_f$ row vector is
\begin{equation}
\tilde{P}_f^T = \begin{pmatrix}
\Re({\rm chan1~prediction~FFT}) \\
\Im({\rm chan1~prediction~FFT}) \\
\Re({\rm chan2~prediction~FFT}) \\
\Im({\rm chan2~prediction~FFT}) \\
\vdots \\
\Re({\rm chanN~prediction~FFT}) \\
\Im({\rm chanN~prediction~FFT}) \\
\end{pmatrix}.
\label{eq:dvecdef}
\end{equation}
Similarly, we can redefine terms in Equation~\ref{eq:ptodata} to accommodate multiple channels. 
First we define an amplitude vector $\vec{a}$ that contains all  amplitudes ($NM$ amplitudes in total):

\begin{equation}
\vec{a}  = (a_{11}, a_{12}, ...~ a_{1M}, a_{21}, a_{22}, ...~ a_{2M}, ...~ a_{N1}, a_{N2}, ...~ a_{NM})
\label{eq:vecp}
\end{equation}

We define ${\bf \tilde{T}}_f$ to be a matrix of size $(NM \times 2N)$ that holds the Fourier transform (real and imaginary parts) of the templates, so that $\vec{a} \cdot {\bf \tilde{T}}_f$ is a $1 \times 2N$ vector that contains the predicted values for the real and imaginary parts of the Fourier amplitudes for each channel, in the same format as Equation~\ref{eq:dvecdef}.  
Each row of ${\bf \tilde{T}}_f$ corresponds to one of the $NM$ amplitudes and contains the real and imaginary Fourier amplitudes at frequency $f$ for each channel resulting from that amplitude.  Given how the amplitudes are defined in terms of channels and templates, only one channel (two columns) of each row is non-zero:

\begin{equation}
{\bf \tilde{T}}_f = \begin{pmatrix}
\begin{matrix}
\Re({\tilde{T}_{11}}) & \Im({\tilde{T}_{11}}) & \\ 
\Re({\tilde{T}_{12}}) & \Im({\tilde{T}_{12}}) & & & & \textbf{\Huge0} & & \\ 
\vdots & \vdots & \\
\Re({\tilde{T}_{1M}}) & \Im({\tilde{T}_{1M}}) &  \\ 
& & \Re({\tilde{T}_{21}}) & \Im({\tilde{T}_{21}}) & \\ 
& & \Re({\tilde{T}_{22}}) & \Im({\tilde{T}_{22}}) & \\ 
& & \vdots & \vdots & \\
& & \Re({\tilde{T}_{2M}}) & \Im({\tilde{T}_{2M}}) & \\ 
 & & & & \ddots \\
& & & & & \Re({\tilde{T}_{N1}}) & \Im({\tilde{T}_{N1}}) & \\ 
& & & & & \Re({\tilde{T}_{N2}}) & \Im({\tilde{T}_{N2}}) & \\ 
& \textbf{\Huge0} & & & & \vdots & \vdots & \\
& & & & & \Re({\tilde{T}_{NM}}) & \Im({\tilde{T}_{NM}}) & \\  
\end{matrix} \\
\end{pmatrix}.
\end{equation}
Note that the templates here are assumed to be those before any time shifts are applied.
%\begin{equation}
%{\bf \tilde{T}}_f = \begin{pmatrix}
%\begin{matrix}{\rm Re}(\tilde{T}_{11}) & \cdots & {\rm Re}(\tilde{T}_{1M}) \\
%{\rm Im}(\tilde{T}_{11}) & \cdots & {\rm Im}(\tilde{T}_{1M})\end{matrix} & & & \bigzero \\
% & \begin{matrix}{\rm Re}(\tilde{T}_{21}) & \cdots & {\rm Re}(\tilde{T}_{2M}) \\
%{\rm Im}(\tilde{T}_{21}) & \cdots & {\rm Im}(\tilde{T}_{2M})\end{matrix} & & \\
% & & \begin{matrix}\ddots \\ \ddots \end{matrix} & \\
%\bigzero & & & \begin{matrix}{\rm Re}(\tilde{T}_{N1}) & \cdots & {\rm Re}(\tilde{T}_{NM}) \\
%{\rm Im}(\tilde{T}_{N1}) & \cdots & {\rm Im}(\tilde{T}_{NM}) \end{matrix} \\
%\end{pmatrix}.
%\end{equation}

Finally, we define a $2N \times 2N$ time shift matrix 
${\bf R}_f(\vec{\delta})$, which is a function of the time delay vector $\vec{\delta}$ containing the time shifts for all channels.  This is a block diagonal matrix of size $2N \times 2N$ where each block is a $2\times 2$ rotation matrix and all other entries outside these $2\times 2$ blocks are zero:
\begin{equation}
\begin{split}
{\bf R}_f(\vec{\delta}) &= {\bf R}_f(\delta_1,\delta_2,\cdots,\delta_N) \\
&= \begin{pmatrix}
\begin{matrix}\phantom{+}\cos{2\pi\delta_1} & \sin{2\pi\delta_1} \\
-\sin{2\pi\delta_1} & \cos{2\pi\delta_1}\end{matrix} & & & \textbf{\Huge0}\\
 & \begin{matrix}\phantom{+}\cos{2\pi\delta_2} & \sin{2\pi\delta_2} \\
-\sin{2\pi\delta_2} & \cos{2\pi\delta_2}\end{matrix} & &  \\
 &  & \begin{matrix} \ddots & \ddots \\
\ddots & \ddots \end{matrix} &\\
\textbf{\Huge0} & & & \begin{matrix}\phantom{+}\cos{2\pi\delta_N} & \sin{2\pi\delta_N} \\
-\sin{2\pi\delta_N} & \cos{2\pi\delta_N}\end{matrix} \\
\end{pmatrix}.
\end{split}
\end{equation}
Each $2\times 2$ block is determined by a $\delta_i$, which is the time delay for that channel. 
${\bf R}_f$ simply applies the phase rotation to each channel.

With the above definitions, we can write the $\chi^2$ from Equation~\ref{eq:chi2f} in matrix form:
\begin{equation}
\begin{split}
\chi^2 &= \sum_f \chi^2_f \\
&= \sum_f \left(\vec{\tilde{D}}_f - \vec{\tilde{P}}_f\right)^T~{\bf V}^{-1}_f~\left(\vec{\tilde{D}}_f - \vec{\tilde{P}}_f\right) \\
&= \sum_f \left(\vec{\tilde{D}}_f - \vec{a}  \cdot {\bf \tilde{T}}_f \cdot {\bf R}_f(\delta_c)\right)^T~{\bf V}^{-1}_{f}~\left(\vec{\tilde{D}}_f - \vec{a}  \cdot {\bf \tilde{T}}_f \cdot {\bf R}_f(\delta_c)\right). 
\end{split}
\label{eq:chi2sum}
\end{equation}
This $\chi^2$ can be analytically  minimized with respect to the amplitudes, with the minimum at
\begin{equation}
\vec{a}^T = \left(\sum_f \left( {\bf \tilde{T}}_f \cdot {\bf R}_f(\delta_c)\right) \cdot {\bf V}^{-1}_f \cdot \left({\bf \tilde{T}}_f \cdot {\bf R}_f(\delta_c)\right)^T
\right)^{-1} \sum_f \left( {\bf \tilde{T}}_f \cdot {\bf R}_f(\delta_c)\right) \cdot {\bf V}^{-1}_f \cdot \vec{\tilde{D}}_f^T .
\label{NxM amp eq}
\end{equation}

\section{Applications in SuperCDMS}
\label{Applications in SuperCDMS}
    
    SuperCDMS employs cryogenic detectors (cylindrical Si or Ge crystals, approximately 10 cm in diameter and $\sim$3 cm thick) for direct detection of dark matter \cite{Rau_2012,  PhysRevLett.112.241302}. % <-- generic SuperCDMS refs here
    Quantized vibrations, called phonons, produced by an incoming particle scattering off an electron or nucleus in the crystal are detected as the primary source of information about the interaction energy.
    %An incoming particle scattering off an electron or nucleus in the target material of the detector (a Si or Ge crystal approximately 10~cm in diameter and $\sim$3 cm thick) produces quantized vibrations, called phonons, and ionization. Although both ionization and phonon signals are utilized in some detectors, the primary detection mode is via phonon signals. 
    Phonon signals are collected bolometrically in multiple sensors on the top and bottom of the detectors, using TESs to convert small thermal excursions into measurable currents which are later amplified and then digitized. 
    
\begin{figure}[!h]
        \centering
        \includegraphics[width = 1\linewidth]{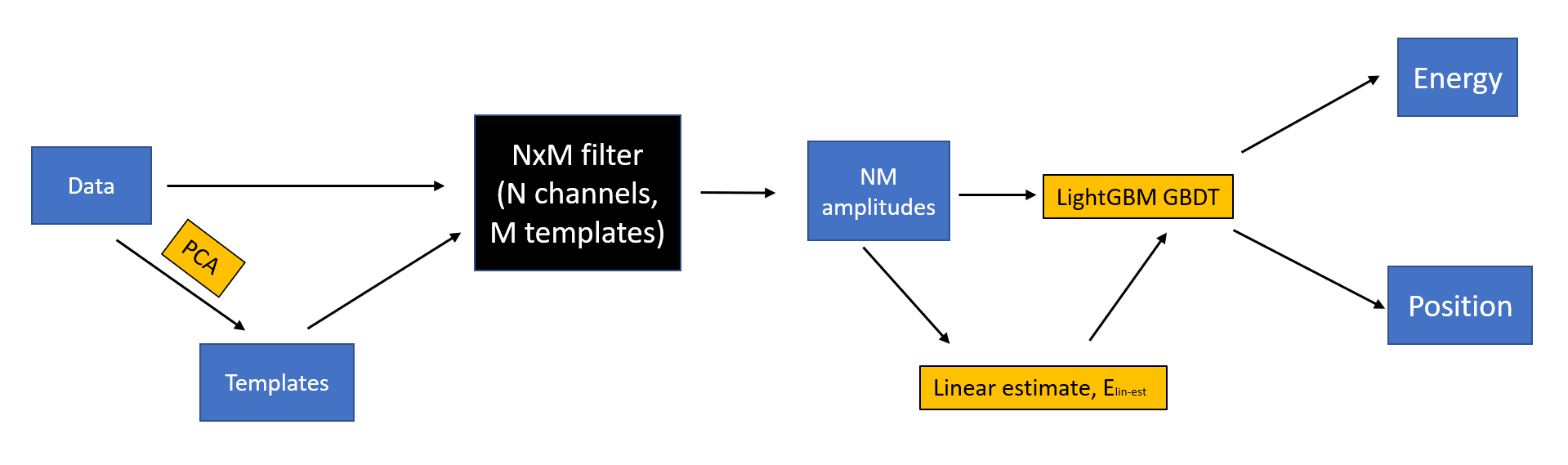}
        \caption{A flowchart showing the N$\times$M filter data processing pipeline. We start from the raw data and construct templates using principal component analysis (PCA). Then we perform N$\times$M processing on the data with the normalized PCA templates to get a total of $NM$ output amplitudes. The $NM$ amplitudes are then used to obtain a linear estimator $E_\mathrm{lin-est}$ which is used to construct labels for GBDT training. The GBDT combines the $NM$ amplitudes to produce a gain correction called $G_\textrm{BDT}$, which is applied to get the final position corrected energy. If the true event positions are known for the training data, one can also reconstruct true position from the N$\times$M amplitudes.}
        \label{fig:NxM pipeline}
    \end{figure}

    %The digitized waveforms contain valuable information about the energy and position of the event in their shapes and amplitudes. 
    Event reconstruction via optimal filters described in the previous sections is our standard method to extract these quantities from a pulse occurring in the waveforms. Many factors affect our reconstruction ability, several of which were described in Section~\ref{Introduction}. The rest of this paper discusses how to apply the N$\times$M filter introduced in Section~\ref{NxM filter} within the context of SuperCDMS.  Our ultimate analysis goal is to obtain a position-corrected energy estimator.  
    The discussion covers the topics of template generation for the N$\times$M filter and using these templates to fit for a total of $NM$ amplitudes as given by Equation~\ref{NxM amp eq}.
    %and interpreting these fitted amplitudes. 
    We will also present the use of machine learning techniques to remap the amplitudes to meaningful physical quantities of interest: energy and position. The overall data processing pipeline is shown in Figure~\ref{fig:NxM pipeline} with more details in the following sub-sections.

    \subsection{Template generation for the N$\times$M filter}
    \label{sec:template_gen}

    The N$\times$M filter allows us to fit multiple ($M$) shapes to each channel (see Equation~\ref{NxM amp eq}). There are two aspects to template generation that require consideration. The first is how one ought to generate the templates. The templates generated must (a) be reasonably independent of detector noise, and (b) capture the variation in pulse shape across the detector. These requirements help prevent unwanted biasing of the filter outputs by ensuring that the templates can accurately represent pulse shapes. 
    
    Furthermore, the template shapes carry useful information about the location of events in the detector. Charge and phonon transport processes result in primary and secondary populations of phonons arriving at the sensors at different times\footnote{For the purposes of time alignment, all templates are constructed with a common starting time, representing the arrival time of the earliest phonons.}.  Prompt phonons preserve information about the location of the original energy deposit in the crystal and contribute a fast component to the pulse shape. Position reconstruction in SuperCDMS detectors is achieved by translating the resulting pulse-shape differences into quantities that correlate with position. In the past, SuperCDMS has used a type of optimal filter called the two-template optimal filter (1$\times$2 OF) which effectively separated the energy and position information in a pulse into amplitudes derived from slow and fast templates, respectively \cite{thakurthesis,pepinthesis}. The amplitude of the fast component primarily provided position information, while the amplitude of the slow template provided an energy estimate.  For this fitter, the slow template was constructed from the average pulse shape, while the fast template was calculated as the average of residuals between many measured pulses and the slow template after inverting those residuals with negative peaks. 
    
    The N$\times$M filter generalizes and systematizes this approach, and so ought to be able to capture the energy and position information more effectively than other methods. 
    %capture not only the energy-dependent component of the pulse shape but also the position-dependent residuals more effectively than other methods.  
    The N$\times$M filter also addresses the issue of correlated noise between channels and the fact that in newer detector designs the pulse shapes cannot be so cleanly separated into a fast and a slow component.
    
    The second aspect of template generation concerns the number of templates. In principle, one can imagine fitting a very large number of templates to a single pulse to obtain a near-perfect match to the measured shape. Such a fit would be limited by computational resources and possibly by degeneracies among the fit parameters. Furthermore, physical interpretation of the results may not be straightforward when using a large number of empirically derived templates.

    With both of these aspects in mind, we present the use of principal component analysis (PCA) \cite{Pearson01111901} to derive effective templates for an N$\times$M fit. The advantage of this approach is two-fold: (a) it presents an elegant mathematical structure that extracts the most important features in the data, and (b) it provides a convenient way to determine the number of \textit{useful} templates. PCA enables identification of the most significant dimensions in a multi-dimensional data set (often termed ``feature identification''). In addition, it provides a framework to achieve dimensionality reduction by decomposing a data set into components (``feature reduction''). We exploit PCA feature identification to obtain templates and feature reduction to restrict the number of templates used in the N$\times$M filter.

    Principal component analysis starts with a data set of $n$ events, each having $p$ numbers associated with it.  A basis vector in this $p$-dimensional parameter space is derived that maximizes the variance of the projection of the events onto this vector.  This procedure is equivalent to finding the direction in the parameter space that explains the largest amount of variance in the data.  After this first principal component is found, its projection is subtracted from each event to remove the variance due to that component, and then the procedure is repeated to find the next basis vector which maximizes the remaining variance in the data.  This process is continued iteratively until a full set of $p$ basis vectors is found.  The procedure is equivalent to finding the eigenvalues of the covariance matrix of the data.  The eigenvectors are called the principal components (PCs), and their associated eigenvalues, arranged in descending order, identify how much of the overall variance in the data can be associated to each component.  In this paper, we will refer to the first component (the one with maximum information or variance) as component 0, the second component as component 1 and so on.

% I rewrote this discuss to be much less mathematical.  See above paragraph.
%    {\color{blue}For a data set consisting of $n$ events, each having $p$ numbers associated with it, PCA begins by constructing a  data matrix, $X$, of size $n\times p$, whose columns represent each of $p$ features with each row containing the observed values of those features for one event. PCA determines an orthonormal set of $p$ $p$-dimensional vectors, $A$, which maximize the variance of the linear combination $XA$.}\footnote{\color{red} Scott: I don't understand the math.  What does it mean to maximize the variance of the linear combination $XA$?  Isn't $XA$ a matrix of size $n\times p$?  So what are we maximizing?  I think the wording is not correct.  The wikipedia article is clearer on this.  Also, what is the relationship between $S$ and $X$ and $A$?  I think this paragraph needs to be rewritten to be a clearer explanation of PCA.} If $S$ is a matrix representing the variance in the data, then it can be shown that the eigenvalues, $\lambda$, of $S$ represent the variance in the data, with max$(\lambda)$ giving the largest variance and the corresponding eigenvector pointing in the direction of maximum variance \cite{cite}. The eigenvectors of $S$ are termed Principal Components (PCs) of the data, often arranged in descending order of the eigenvalues. In this paper, we will refer to the first component (component with maximum information or variance) as component 0, the second component as component 1 and so on.

%    SuperCDMS SNOLAB data 
    Data from the new detectors built for SuperCDMS SNOLAB
    consist of digitized waveforms (called traces) sampled at a frequency of 625~kHz (2.5~MHz) for phonon (charge) channels. Each time bin of the waveform contributes information about the position, energy, and noise of an event. A waveform containing $p$ time bins therefore has $p$ features. To determine templates for the N$\times$M filter, we apply PCA to a sample of $n$ traces each with a trace length of $p$ samples, aligned with respect to the start point of the pulse in the trace. The start point is typically configured to occur near the midpoint of the trace. Alignment of traces is necessary to avoid capturing artifacts in the PCs due to small differences in pulse start times. To ensure that the PCs capture pulse-shape variation across the full detector, we compute a common set of templates (to be used for all detector channels) from digitized events that include traces from all detector channels. We normalize the templates by the maximum height of their respective PCs. 
    
\begin{figure}[!h]
        \centering
        \begin{subfigure}{1\textwidth}
            \centering
            \includegraphics[width=16cm,height =8cm]{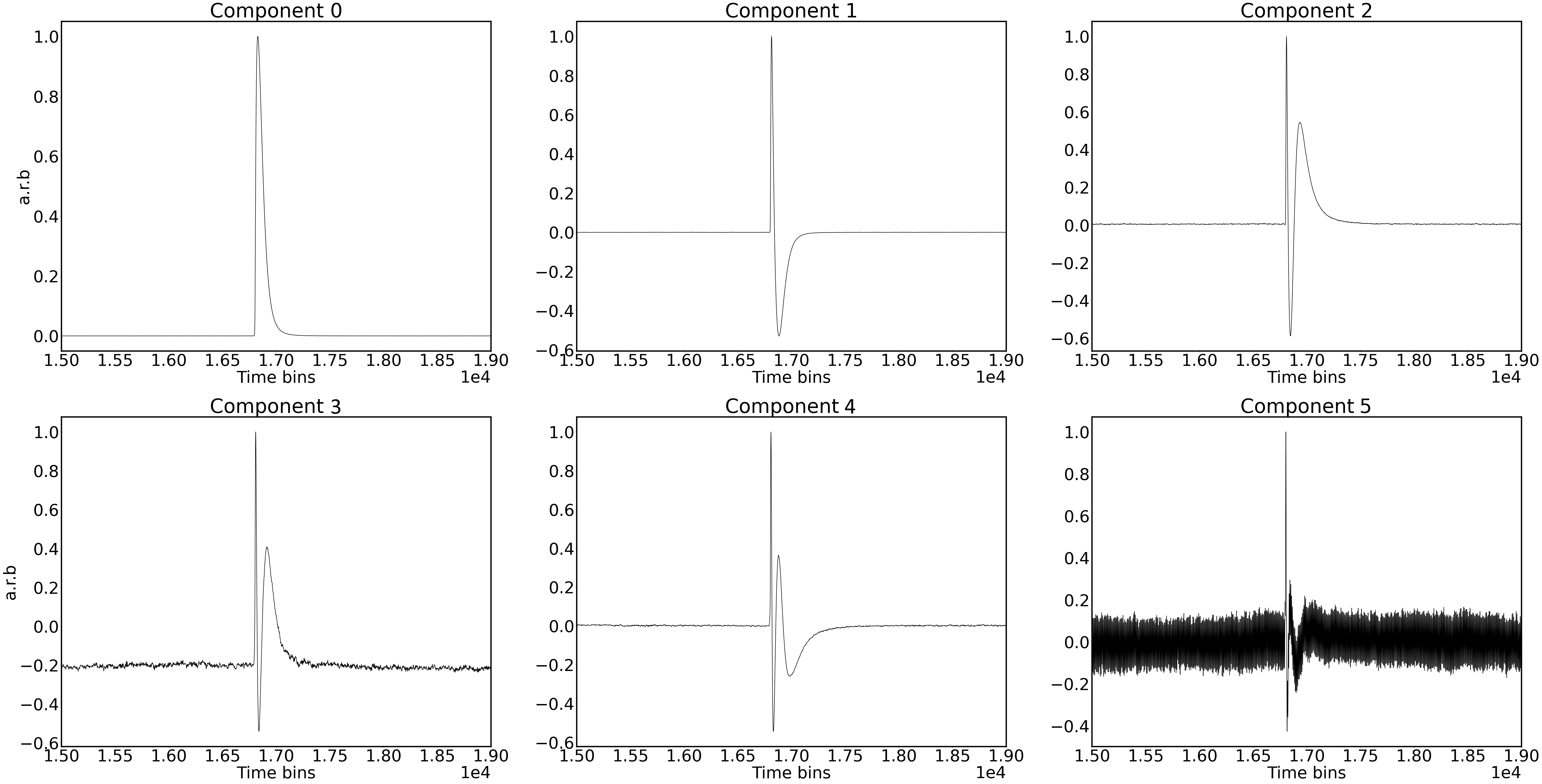}
            \caption{}
            \label{fig:sfig1}
        \end{subfigure}%
        \\
        \begin{subfigure}{1\textwidth}
            \centering
            \includegraphics[width=0.4\linewidth]{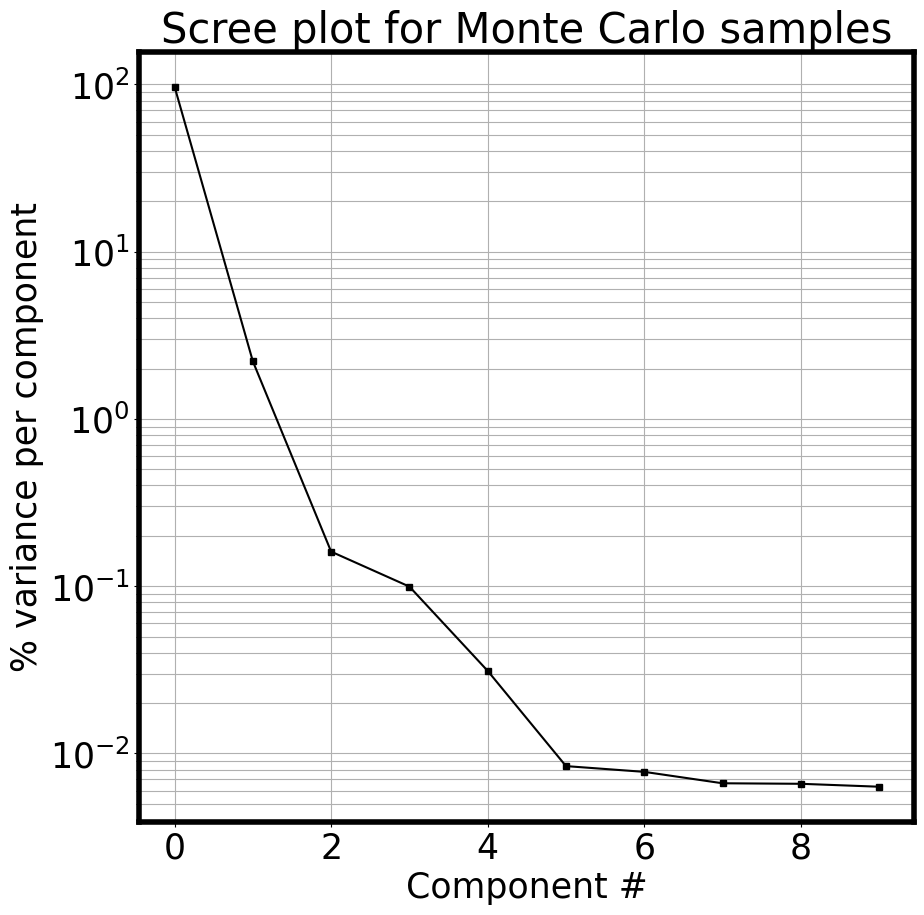}
            \caption{}
            \label{fig:sfig2}
        \end{subfigure}
        \caption{ Results from applying PCA to the simulated data set described in Section \ref{With GEANT4 simulations}. Figure (\ref{fig:sfig1}) shows the first six principal components. Figure (\ref{fig:sfig2}) shows the corresponding scree plot up to ten components. The scree plot shows that information stored in the fourth or higher templates is less than 0.1\% and hence for all practical purposes essentially negligible. %.The scree plot shows that most of the information is stored in the first 3 components. 
        Noise contributions are more evident in higher components.
%        In accordance, the amount of noise in the templates increases with component number. 
 In this example, one would choose to perform a fit with the first 3 components only because the subsequent components add information at the sub-percent level while increasing the computational complexity of the fit.}
        \label{fig:DMC PCA}
    \end{figure}

    PCA produces $p$ PCs for a data set with $p$ features. $p$ corresponds to the trace length in our case and it can be as large as 32768 (131072) samples for phonon (charge) traces corresponding to a 52~ms trace. It is therefore important to determine the optimal balance between number of templates and amount of useful information in the templates. We produce a scree plot \cite{screeplot} which displays the amount of information in each component as a function of component number. The $y$ axis is the fractional contribution of the eigenvalue or variance of that component to the total variance in the data set. Examples of PCs and the corresponding scree plot for a simulated data set (which is described in Section~\ref{With GEANT4 simulations}) are shown in Figure~\ref{fig:DMC PCA}. We normalize each PC in Figure~\ref{fig:DMC PCA} so its maximum height is equal to 1. We perform a qualitative assessment of the PCs and the scree plot to determine a final template selection. 
    
    %Our ultimate analysis goal is to combine the amplitudes across the different channels to obtain a position-corrected energy estimator. 
   We aim to combine the amplitudes across different channels to obtain a position-corrected energy estimator. 
    We achieve this via a two-step process which first calculates a linear estimator from the N$\times$M amplitudes and then corrects for higher-order effects with supervised machine learning. This procedure is described in Sections~\ref{linear energy reconstruction} and \ref{Machine learning reconstruction}.
   % in the following sections.
    
    \subsection{Energy reconstruction with linear estimator}
    \label{linear energy reconstruction}

    The $M$ normalized PCs obtained from the procedure described in Section~\ref{sec:template_gen} are used to generate amplitudes for each of the $N$ channels using the N$\times$M filter. In all, the N$\times$M filter produces a total of $NM$ amplitudes which represent the relative contribution of each PC. Each amplitude contains information about both the energy and position of the event, with each channel's component 0 containing most of the energy information. A weighted sum of the channel amplitudes provides a reasonable, first-order estimate of the reconstructed energy: 
    \begin{equation}
        E_\mathrm{lin-est} = \sum_{i=0}^{NM-1} P_i \cdot\alpha_i,
        \label{E_lin_est}
    \end{equation}
    where $P_i$ is the $i^{th}$ amplitude from the N$\times$M fit and $\alpha_i$ is the corresponding weight for that component.  (Note that this sum is over both channels and principal components.)
    This estimate corrects for linear dependencies, such as relative calibration\footnote{Relative calibration factors are weights applied to each channel to account for unequal TES response owing to %channel noise, 
    differences in bias points, manufacturing tolerance, etc.} of the channels, by appropriately weighting the contribution of each component as follows. The weights, $\alpha_i$, are obtained by minimizing the following $\chi^2$ summed over a set of training events:
    \begin{equation}
        \chi^2 = \sum_j\left[ E_\mathrm{true}^{j} - \sum_i \left(P^j_i\cdot{\alpha_i}\right)\right]^2,
        \label{chisq}
    \end{equation}
    where $E_\mathrm{true}^j$ is the true energy for event $j$ (for example, simulated events of known energy or events selected in a calibration peak). The subscript $i$ and superscript $j$ on $P_i^j$ denote the $i^{th}$ amplitude component for the $j^{th}$ event. Equation~\ref{chisq} can be analytically minimized to give the weight of the $i^{th}$ amplitude component:
    %\begin{equation}
    %\label{alpha}
    %    \vec{\alpha} = \left( P^TV^{-1}P\right)^{-1}P^TV^{-1} \cdot E_\mathrm{true}.
    %\end{equation}
    \begin{equation}
    \label{alpha}
        {\alpha_i} = \frac{\sum_j P^j_i \cdot E_\mathrm{true}^j}{\sum_j \left(P^j_i\right)^2}.
    \end{equation}
       These weights can be calculated provided we have a data set with events whose true energies are known. Calibration data sets with readily identifiable features can be used for this purpose.  Generally speaking we use a mono-energetic set of events, such as from a calibration peak, to determine a set of weights.

When applied to a set of mono-energetic events, Equations~\ref{E_lin_est} and \ref{chisq} implicitly include a partial correction for position dependence in the estimated energy.  Position dependence in the fitted amplitudes will tend to broaden the energy resolution.  By minimizing the $\chi^2$ in Equation~\ref{chisq}, we produce a set of weights that yields a linear estimator that minimizes the energy resolution, including an effective linear correction for any position dependence of the amplitudes.  These weights will also reflect the noise levels in the detector, since amplitudes that are particularly affected by noise will be de-weighted as a result of this procedure.
%The estimate defined in Equation~\ref{E_lin_est} is not only the linear estimator with minimal statistical deviation from the true energy, $E_\mathrm{true}$, but also implicitly includes a linear correction for the position dependence of the reconstructed energy.  

%It is important to point out that the weights are energy dependent: 
    It follows that the optimal weights (those which minimize Equation~\ref{chisq}) are energy dependent: 
    ${\alpha(E)} = \alpha_0 + \beta(E)$, where $\alpha_0$ is a constant and 
    $\beta(E)$ is some function of energy.  This energy dependence results from the fact that the relative contributions of noise and position dependence vary with energy, with position-dependent effects dominating over noise at higher energies.   The $\beta(E)$ term accounts for these position effects.  Because the linear weights depend on both the noise levels in individual channels and on position dependence of the pulse amplitudes and shapes, which are not known {\em a priori}, the weights to be applied for any given event must be derived from data. 
    We have devised an iterative method to approximate the weights from data using calibration peaks with known true energies (such as mono-energetic signatures resulting from decays of $^{71}$Ge). A preliminary energy estimate is calculated by setting the weights to one for every event. This is sufficient to allow us to identify peaks in the energy spectrum.  Then, using events from the calibration peaks, we estimate the weights at those energies. We assume that the same weights can be applied to events within 5$\sigma$ of the calibration peak. 
    %When estimating the weights to be used for events outside of the peaks, a preliminary energy estimate is done using weights of one.  
    The preliminary energy estimates for events outside the peaks are then used to linearly interpolate the weights,
    %This preliminary energy estimate is then used to linearly interpolate the weights, 
    and those interpolated weights are then used in Equation~\ref{E_lin_est} to generate an improved energy estimate. Events whose preliminary energy estimates fall above or below the 5$\sigma$ region of the first and last calibration peaks, respectively, are assigned the same weights as the nearest calibration peak. %The final energy estimator is calculated using the interpolated weights for each event. 
%   This introduces systematic uncertainty from the linear interpolation of the weights based on a small number of known points, which we 
 The use of only a small number of known energy points for the linear interpolation of the weights introduces systematic uncertainty, which we acknowledge as a limitation of the technique that can be studied in more detail with future simulation datasets. Note that we have introduced an additional requirement -- the need for multiple calibration peaks in the data set -- to obtain optimal energy resolution from the N$\times$M filter.

    \subsection{Machine learning reconstruction of energy and position from $N \times M$ amplitudes}
    \label{Machine learning reconstruction}
    It may be possible to further improve the resolution performance relative to Equation~\ref{E_lin_est}. The linear estimate is corrected for correlated noise implicitly with the non-diagonal covariance matrix in Equation 22, and the resolution broadening induced by position dependence is corrected to first order as described above. However, the reconstructed energies can manifest non-linear dependencies on the event location inside the crystal. The nature of these higher-order terms is nearly impossible to determine analytically even with the availability of accurate simulations whose data are indistinguishable from experimental data. In cases such as this, where the model that maps a set of variables to an output is unknown, one can use machine learning to evaluate an empirical model. We  use Gradient Boosted Decision Trees (GBDTs) \cite{gbdt2} from the LightGBM package \cite{lightgbm} to estimate non-linear corrections to energy from position dependence. The use of PCA templates constructed in the manner described in Section~\ref{sec:template_gen} allows us to capture position information in the amplitudes.  If these amplitudes are used as input variables, the GBDT combines the amplitudes in an optimal fashion which can improve the resolution of a chosen calibration peak.

    Although decision trees are well known in the context of classification problems, here we use them to perform regression tasks. Let us consider the case of a one-dimensional mapping of a variable $X$ to another variable $Y$. Assume that the exact nature of the relationship between $X$ and $Y$ is unknown and therefore needs to be calculated empirically.  GBDTs work by finding the ideal way to split the given data in $X$ into categories such that each category minimizes the prediction error for $Y$. Mean Squared Error is the commonly used cost function to evaluate the efficacy of each split. 
    %In multi-dimensional data, i.e. a mapping of a set, $\{X\}$, to an output, $Y$, the splitting is performed in a multi-dimensional space by optimizing the cost function in all $X$. 
    In multi-dimensional data, i.e. a mapping of a set, $\{X\}$, to an output, $Y$, the GBDT algorithm attempts to divide the multi-dimensional space so as to minimize the cost function summed over events in the training set.    
    An individual tree that performs such a regression does not necessarily produce the best results and is often termed a weak predictor. Therefore, the method is expanded to an ensemble of decision trees, each of which on its own is a weak predictor, but jointly act as a strong 
    predictor~\cite{gbdt4}. Each successive tree learns from the errors of its predecessors via gradient descent. GBDTs have the general advantage of being more interpretable compared to neural networks, even though visualizing a tree for high-dimensional data may not be straightforward.

     LightGBM GBDT is effective on comparatively small data sets \cite{gbdt1}, which makes it a suitable algorithm to employ in a rare event search experiment such as SuperCDMS. We use LightGBM GBDT in the following way. The $NM$ amplitudes from the N$\times$M filter are rescaled to be of similar order of magnitude to ensure that no single amplitude is unfairly valued by the algorithm. The rescaled amplitudes serve as the features that the GBDT trains on. The goal is to map the $NM$ amplitudes to an energy correction factor. We call this energy correction factor $G_\textrm{BDT}$, which is meant to supplement the typical linear position gain corrections employed in SuperCDMS \cite{PhysRevD.97.022002}. $G_\textrm{BDT}$ is defined as the fractional deviation of true energy from $E_\mathrm{lin-est}$:

    \begin{equation}
        \label{BDTGain}
        G_\textrm{BDT} = \frac{E_\mathrm{true}-E_\mathrm{lin-est}}{E_\mathrm{lin-est}}.
    \end{equation}
   We assume that this correction factor is independent of $E_\mathrm{true}$, corresponding to the case that the correction results from non-uniformities in phonon collection efficiency that are independent of the true energy.  This assumption allows us to generalize a sample trained on a single calibration peak to other energies.  Use of training samples with a larger range of energies, which are currently not available, would allow us to relax this assumption. %Use of larger training samples may allow this assumption to be relaxed in the future.
Among the hyperparameters that define the structure of the GBDT are the number of trees in the ensemble, the maximum number of leaves in the trees, the depth of the tree, and the minimum number of samples in a child node for the next split. A grid search on a validation sample was employed to find the optimal number of trees, leaves and depth.

    Calibration peaks provide events with known energies which can be used to derive $G_\textrm{BDT}$-trained models. Obtaining position labels is not possible for real data sets, as we do not know the true positions of events in the detector. However, SuperCDMS simulations based on GEANT4  can produce both position- and energy-labeled data sets. Although it may not be fully representative, we can use the simulated data set (described in Section~\ref{With GEANT4 simulations})  to demonstrate a proof of concept. In real data, the training is performed on events from one calibration peak, validated on a statistically independent sample of events from the same calibration peak, and then tested on a completely different calibration peak to evaluate generalizability across the full energy spectrum. The final corrected energy for a new data point after training is given by
    
    \begin{equation}
        \label{BDT corrcted energy}
        E_{{\rm N}\times {\rm M}} = E_\mathrm{lin-est}\cdot\left( 1 + G_\textrm{BDT}\right).
    \end{equation}

    %We emphasize that there is no ``one model fits all" solution for our data. 
    The fundamental assumption in training any machine learning model is that the training and test data sets are independently derived from the same underlying distribution. In our stable detector operation modes, such as in a dark matter data set, we fix important physics parameters such as the TES bias points and the detector voltage. Although the noise environment could potentially change with time, 
    %this is optimally deweighted within the mathematical framework of the N$\times$M filter. 
    the covariances and weights used in the N$\times$M filter can readily be updated to optimally handle the noise.   
    However, the operational bias points and pulse shapes of each detector vary based on the particulars of its design, material, and fabrication variances, which in turn manifest themselves in our input $NM$ amplitudes. Therefore, a separate GBDT needs to be trained for each detector; however, once trained, the same model can be applied to all data sets acquired under the same operating conditions for the same detector. Moreover, the hyperparameters that define the structure of each GBDT need to be tuned separately for each detector. 
    
    %Sometimes a regularization term was also considered to prevent overfitting. 
    
    For most real data sets, our inputs are  the $NM$ amplitudes and the output to be trained is the value of $G_\textrm{BDT}$. In the case of simulated data, we know the true position of the events in addition to their energies, which allows us to map the amplitudes to position as well. This is not the case if training on real data whose position labels are unknown. For simulated data, one can slightly alter the model to map the $NM$ features to a $G_\textrm{BDT}$ value and three position parameters corresponding to the three geometric axes of the detector. However, a model trained on simulated data is not guaranteed to produce best results on real data unless the simulation parameters are tuned for that detector.

    The pipeline in Figure~\ref{fig:NxM pipeline} was tested on data drawn from simple analytic models and on simulated events and finally applied to real data from CDMSlite \cite{PhysRevLett.112.041302}, a study done with a previous generation of SuperCDMS detectors operated in the Soudan mine. 
    %The toy models were too simplistic to be worth elaboration in this paper. 
    In the following sections, we describe results from applying the data processing pipeline to GEANT4 simulations and CDMSlite data.

\section{Validation of the $N \times M$ analysis framework}
\label{Validation}
    
    In Section~\ref{With GEANT4 simulations} we present results from applying the N$\times$M data processing pipeline to simulated events, followed by results from applying the pipeline to CDMSlite data in Section~\ref{ With CDMSlite data}.
    
    \subsection{Validation with GEANT4 simulations}
    \label{With GEANT4 simulations}
 
 We used the SuperCDMS Detector Monte Carlo (DMC) framework to simulate 10,000 mono-energetic (0.5~keV) events distributed uniformly throughout the volume of a SuperCDMS SNOLAB-style High Voltage (HV) detector operated at 100 V ($\pm$50 V on either face of the detector) \cite{cdms-snomass}. HV detectors produce an amplified phonon signal proportional to the applied voltage ($V$) via the Neganov-Trofimov-Luke (NTL) effect \cite{ntl1,ntl2,ntl3}, 
 \begin{equation}
     \label{eq:NTL eq}
     E_{t} = E_{r} + n_{eh}eV.
 \end{equation}
 Here, $E_r$ is the original recoil energy of an electron for a measured energy $E_{t}$ and $n_{eh} = E_r\cdot Y(E_r)/\epsilon$ is the number of electron-hole pairs produced in the interaction\footnote{$\epsilon$ is the ionization energy. $Y(E_r) \equiv 1$ for electron interactions and is an energy-dependent factor $<1$ for nuclear recoils.}. HV detectors are projected to achieve excellent sensitivity to low mass WIMPs. They contain 12 phonon channels and do not have any direct charge readout. The simulation uses GEANT4 \cite{Geant4} to generate particle hits and G4CMP \cite{g4cmp} to simulate phonon and charge transport in the crystal, and combines these with the detector response to produce simulated events. The noise PSD from a real data set is used to generate noise traces to add to the simulated pulses. The noise described by a PSD does not account properly for correlations between channels and hence the noise data used in this example is not correlated. The DMC runs an extensive pipeline which produces raw traces in the data format used by the experiment.

        \subsubsection{Energy Reconstruction}
        \label{DMC energy reconstruction}
        
\begin{figure}[!h]
        \centering
        \includegraphics[scale=0.6]{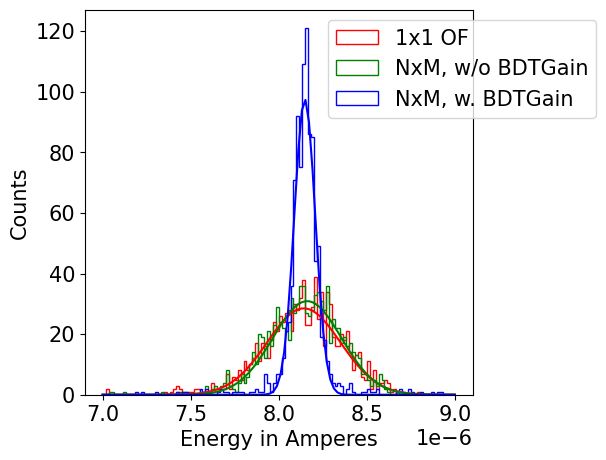}
        \caption{A comparison of the reconstructed energy from different estimators for the mono-energetic simulated samples. The estimators being compared are the simple 1$\times$1 OF (red), the N$\times$M filter without any position correction (green) and the N$\times$M filter with position correction (blue).}
        \label{fig:DMC spectrum}
    \end{figure}

        The raw traces produced from the simulation are run through the full processing pipeline shown in Figure~\ref{fig:NxM pipeline}.  Following the pipeline, we first determine the N$\times$M templates. Figure~\ref{fig:DMC PCA} shows that only the first three PCA components contain significant information about the pulse shape. Therefore, we perform a 12 channel $\times$ 3 template fit to obtain 36 amplitudes. The N$\times$M amplitudes produced are first used to construct $E_\mathrm{lin-est}$, and then the data set is split 90/10 to perform supervised learning. 
       A training sample consisting of 90\% of the simulated events is used to train the GBDT, with the remaining 10\% used to assess the performance of the model in terms of the width of the reconstructed peak. Note that unlike a real data set, here we are restricted to a single mono-energetic peak for training and testing. The resolution of the peak on the test sample is estimated by fitting a Gaussian to a histogram of the reconstructed energies. A comparison to the width of the peak obtained from the simple 1$\times$1 OF described in Section \ref{Introduction} is shown in Table \ref{DMC resolution table}. We find that the N$\times$M filter without GBDT corrections is statistically equivalent to the simplest OF applied to the sum of phonon channels. This observation is attributed to the fact that the events used in the simulation do not have any added correlated noise. Hence the statistical treatment of correlated noise in Equation~\ref{NxM amp eq} does not provide significant enhancements in energy resolution compared to the simple 1$\times$1 OF. While E$_\textrm{lin-est}$ corrects for linear effects of position dependence, this appears to be a sub-dominant contribution in this particular data set. Subsequent application of $G_\textrm{BDT}$ using Equation~\ref{BDT corrcted energy} provides a factor of $\sim$3.3 improvement in resolution of the test data showing that position effects in these data are predominantly non-linear in nature. A comparison of the spectrum in each case is shown in Figure~\ref{fig:DMC spectrum}.

    \begin{table}[h]
        \centering
        \begin{tabular}{||c|c||}
            \hline\hline
            Estimator & Peak resolution in \% \\
            \hline\hline
            1$\times$1 OF & 2.56\% $\pm$ 0.14\% \\
            \hline
            N$\times$M OF without position correction (E$_\mathrm{lin-est}$) & 2.39\% $\pm$ 0.13\%\\
            \hline
            N$\times$M OF with $G_\textrm{BDT}$ position correction (E$_{{\rm N}\times {\rm M}}$) & 0.73\% $\pm$ 0.04\%\\
            \hline\hline
        \end{tabular}
        \caption{Comparison of resolution between N$\times$M with and without position correction and 1$\times$1 OF.}
        \label{DMC resolution table}
    \end{table}

      % Scott start reading here  
        \subsubsection{Position Reconstruction}
        \label{Position Reconstruction}
        
\begin{figure}[!h]
        \centering
        \includegraphics[scale = 0.25]{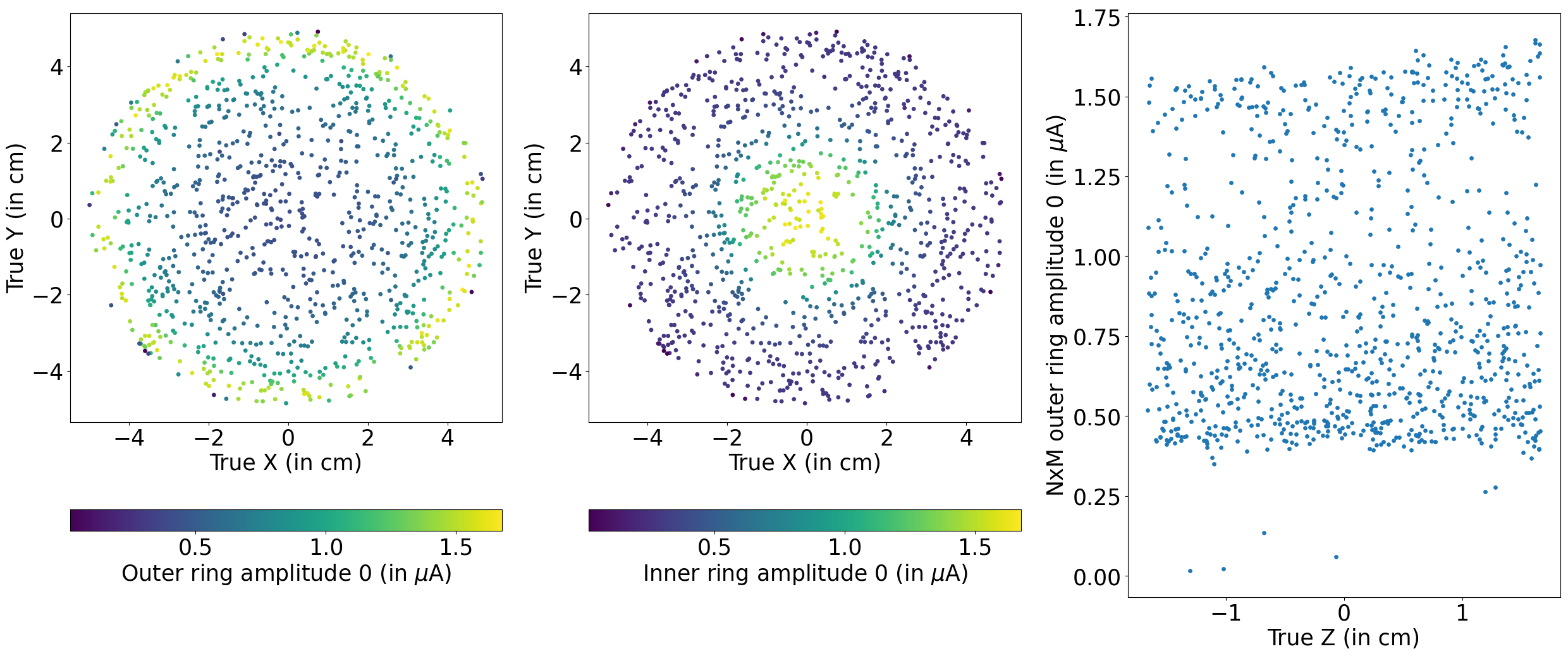}
        \caption{A comparison of the N$\times$M amplitude distribution as a function of true position for the simulated data set. The first two panels show the distribution of N$\times$M amplitude 0 for the outer and inner rings in the X-Y plane of the detector. The third panel shows the N$\times$M amplitude 0 as a function of true Z position for the outer ring channel. While the amplitude is non-linearly correlated to the true X,Y positions, the true Z position shows negligible correlation which is also reflected in the performance of the GBDT.}
        \label{fig: Amps vs position}
    \end{figure}

\begin{figure}
        \centering
        \begin{subfigure}{0.4\textwidth}
            \centering
            \includegraphics[width=2.2in]{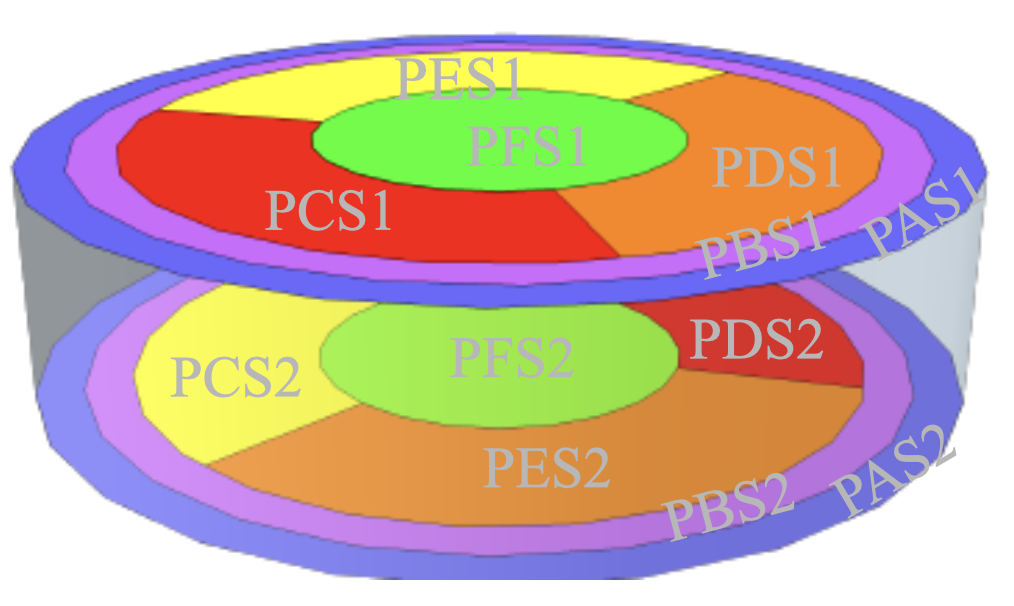}
            \caption{HV detector}
            %\label{fig:sfig_channelmap}
        \end{subfigure}
        \begin{subfigure}{0.5\textwidth}
            \centering
            \includegraphics[width=2in]{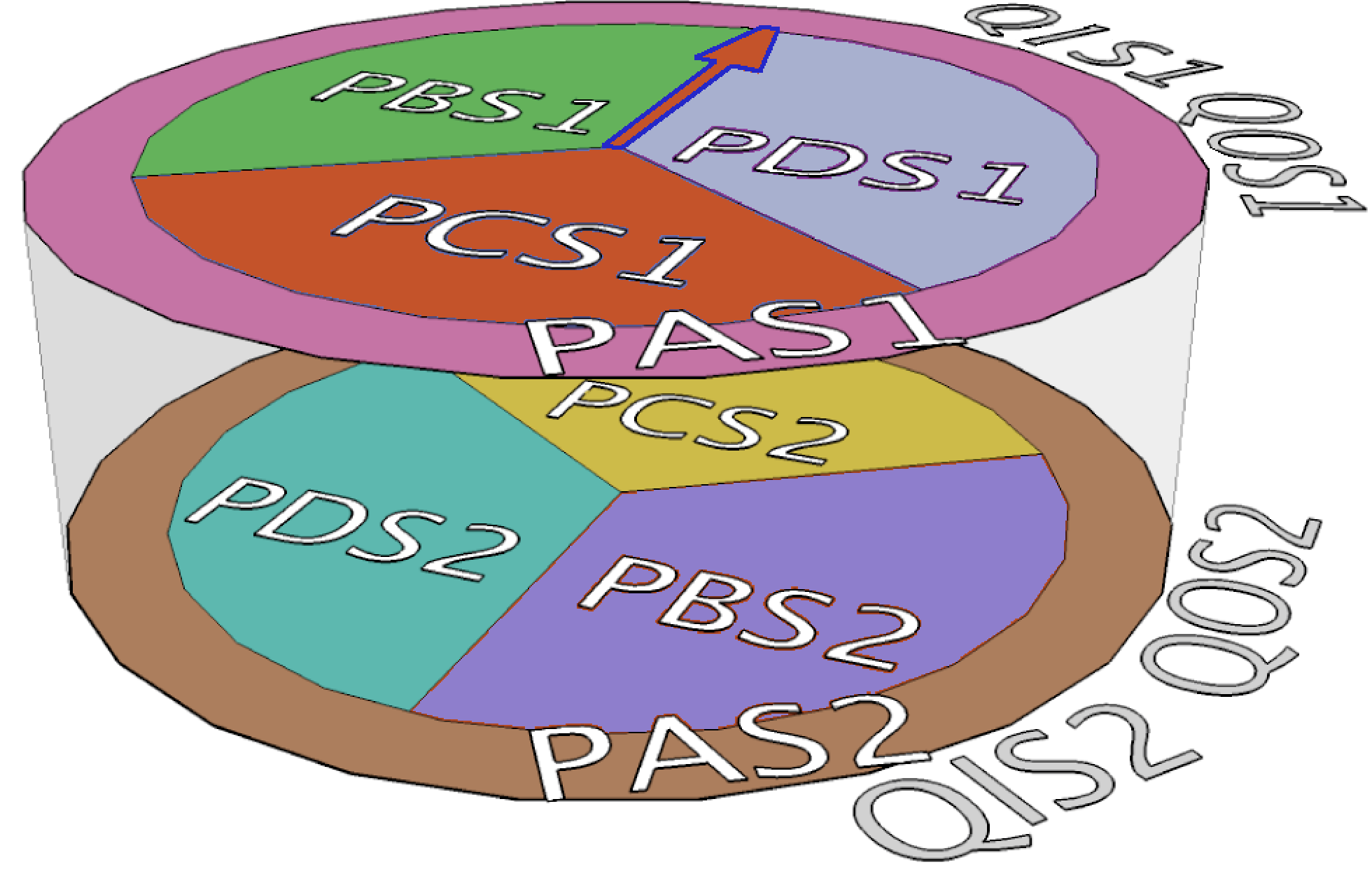}
            \caption{Soudan iZIP detector}
        \end{subfigure}
        
        \caption{The sensors of SuperCDMS detectors are partitioned into channels to enable effective fiducialization of the detector volume. This figure shows the channel map for (a) a SuperCDMS SNOLAB HV detector and, (b) a SuperCDMS Soudan iZIP detector.}    \label{fig:sfig_channelmap}
        
    \end{figure}

        It is straightforward to verify that the N$\times$M amplitudes contain position information.  Figure~\ref{fig: Amps vs position} shows the dependence of some of the  N$\times$M amplitudes on position. A channel map of the simulated detector is shown in Figure~\ref{fig:sfig_channelmap}. In general, an event located closer to a certain channel deposits more energy in that channel compared to others. This effect is clearly visible in Figure~\ref{fig: Amps vs position}. For example, events near the outer ring channel (PAS1) deposit the least energy in the central channel (PFS1) and vice versa (see Figure~\ref{fig:sfig_channelmap}). The high level of non-linear dependence of amplitudes on position is also evident in these plots. Furthermore, the correlation of each component amplitude with true event position is non-linear and different for different channels, which justifies our use of the GBDT to perform a non-linear correction to the N$\times$M energy estimator.

\begin{figure}[!h]
        \centering
            \centering
            \includegraphics[scale = 0.25]{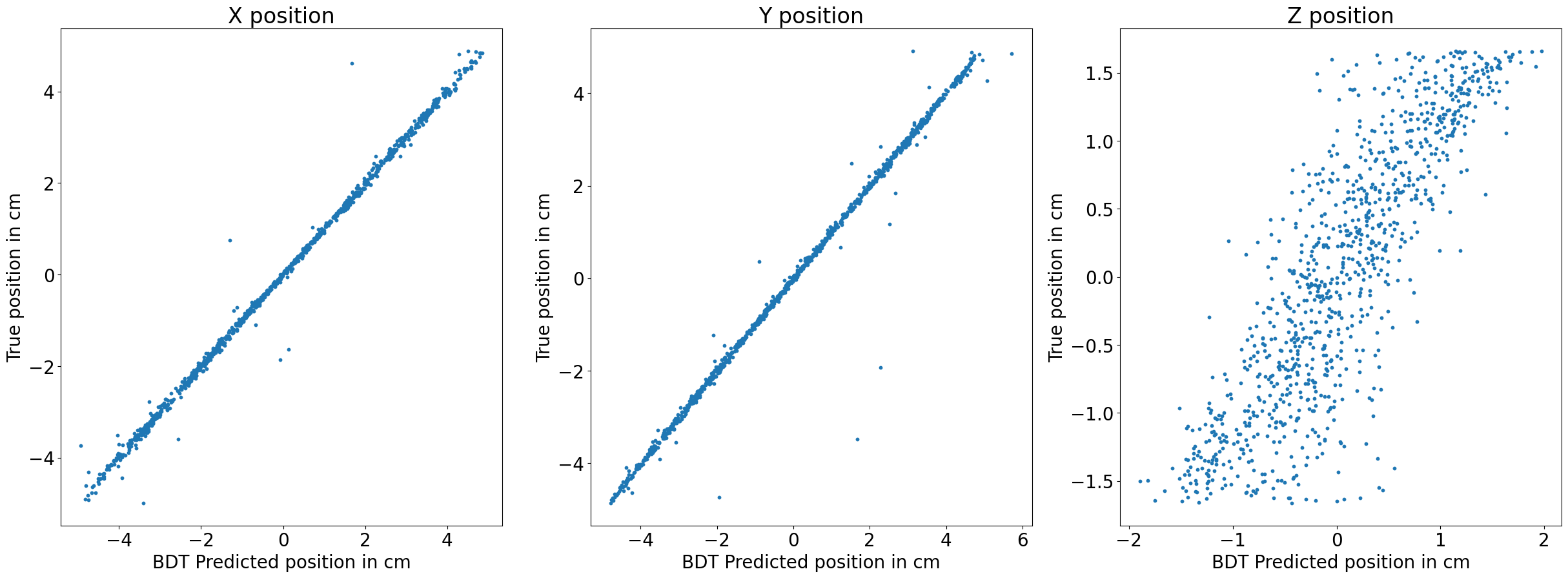}
            \label{fig:sfig_true vs predict}
        
        \caption{A scatter plot of the reconstructed position from the BDT vs. true position in X, Y and Z directions, respectively from left to right. The BDT does not reproduce the Z position very well. The simulated events were distributed uniformly in volume throughout the detector.
        %All three parameters are uniformly distributed denoting homogeneity in events across the detector.
        }
        \label{fig: Predicting position}
    \end{figure}

     An additional advantage of working with simulated data is that, unlike real data, we can train on position labels to reconstruct true position from the N$\times$M amplitudes. We use the same GBDT but supply it with 3 additional labels to reconstruct position: X, Y and Z. We find that the GBDT is able to reconstruct X and Y position with excellent accuracy (within 0.52 and 0.55 mm, respectively) as shown in Figure~\ref{fig: Predicting position}. We also find that Z position reconstruction is poor (prediction error is about 5.49 mm), which is not surprising given the relatively low level of correlation in Figure~\ref{fig: Amps vs position} between the amplitudes and true Z positions. This lack of correlation is expected in HV detectors because the NTL effect rapidly produces additional phonons along a column in the Z direction, which smears out Z-position sensitivity. Inclusion of phonon arrival times from an appropriately defined time delay quantity has been shown to help reduce this effect. Furthermore, the effectiveness of applying a machine learning model trained on simulated data to a real data set remains to be verified. We emphasize once again that at this point in time, the most realistic simulations available for SuperCDMS detectors are not fully representative of real pulses. Additional work is required to extend the application of this position reconstruction approach to SuperCDMS SNOLAB.

    \subsection{Validation with CDMSlite data}
    \label{ With CDMSlite data}

    CDMSlite was an experiment conducted as part of SuperCDMS Soudan. It involved running the conventional SuperCDMS detector at the time, the iZIP, which had both phonon and charge readout, in ``CDMSlite mode" \cite{PhysRevD.97.022002, PhysRevLett.112.041302, PhysRevLett.116.071301,  PhysRevD.99.062001}. The detector was grounded on one side, and a high voltage of up to $V=-75$\,V was applied on the other side to produce NTL amplification. 
    
    Three successful runs with iZIP detectors in this mode inspired the design and development of HV detectors for SuperCDMS SNOLAB \cite{PhysRevD.97.022002, PhysRevLett.116.071301,PhysRevD.99.062001}. We have applied the N$\times$M pipeline in Figure~\ref{fig:NxM pipeline} to data from CDMSlite Run 3 \cite{PhysRevD.99.062001}. CDMSlite Run 3 operated a Ge detector for a period of about 5 months. $^{70}$Ge, when exposed to a neutron source, produces ${}^{71}$Ge, which decays via electron capture with a half life of 11.43 days \cite{PhysRevLett.116.071301, SCHONFELD19981353}. The subsequent x-ray and Auger-electron emissions generate three distinct peaks corresponding to the shell from which the electron was captured: the K-shell peak at 10.37 keV, the L-shell-peak at 1.30 keV and the M-shell peak at 160 eV. These peaks are used for the energy calibration of the detector.

    CDMSlite used the Non-Stationary Filter (NF) described in Section~\ref{Dealing with pulse shape variation} as the final energy estimator for its dark matter analysis, after correcting for various environmental effects (e.g. temperature) with gain factors and a simple correction for position dependence using the amplitude of the fast amplitude from the 1$\times$2 OF (described briefly in Section~\ref{sec:template_gen}) \cite{PhysRevD.97.022002}. The main part of the analysis which differs for the N$\times$M energy estimator is the position correction. In the CDMSlite dark matter analysis, the NF estimator, $E_\textrm{NF}$, was found to correlate with the fast amplitude and therefore a linear correction was devised to decorrelate the two, thus effectively correcting for position effects. This correction took the form

    \begin{equation}
        \label{E_NF position correction}
        E^\mathrm{corrected}_\textrm{NF} = E^\mathrm{uncorrected}_\textrm{NF}\cdot\left(1 + C\cdot [A_\mathrm{fast} - A_\mathrm{fast}^\mathrm{mean}] \right),
    \end{equation}
    where $A_\mathrm{fast}$ denotes the fast amplitude with the superscript ``mean'' denoting the mean fast amplitude obtained from the K-shell events, $C$ is a constant, and the superscript ``uncorrected'' denotes the estimator prior to position correction. The final optimized estimator, $E^\mathrm{corrected}_\textrm{NF}$, was used in the dark matter analysis. Similarly, for comparison in this study, we also construct the 1$\times$2 OF estimator, E$^\mathrm{corrected}_{1\times2}$, after similar corrections are applied.

    Here we attempt to compare our best position-corrected estimator from the N$\times$M, i.e., $G_\textrm{BDT}$-corrected $E_\mathrm{lin-est}$ (E$_{N \times M}$), with two other energy estimators: NF ($E^\mathrm{corrected}_\textrm{NF}$) and 1$\times$2 OF (E$^\mathrm{corrected}_{1\times2}$), used in CDMSlite. The sensors of the Soudan iZIP detectors were sectioned into 8 phonon readout channels, compared to 12 in SuperCDMS SNOLAB detectors (see Figure~\ref{fig:sfig_channelmap}), of which 4 were not read out in the CDMSlite measurements to accommodate application of a large voltage bias on the detector. PCA analysis showed that most of the information was contained in the first 4 PCs. Therefore, we performed a 4$\times$4 fit to the CDMSlite data. The 16 amplitudes produced from this processing were only corrected for environmental effects that showed a correlation with them. They are then combined using Equation~\ref{E_lin_est} with weights estimated from Equation~\ref{alpha}.
    
   The final step is to correct for position dependence. 
    %We do not have position labels to train the GBDT on, hence we will not attempt to reconstruct the absolute position of the events in the detector.
    Since we do not have position labels to train the GBDT, we do not attempt to reconstruct the events' absolute positions.  Energy corrections with $G_\textrm{BDT}$ were calculated with the GBDT. Unlike the simulated data set in Section~\ref{With GEANT4 simulations}, in this case we have multiple independent energy peaks to train, validate, and test our machine learning model. The smaller number of features in this data set (16 amplitudes) allows us to obtain reasonable results even with only thousands of training data points. We select events in the Ge K-shell activation peak in the spectrum to train the GBDT. The events in the peak are selected following the same procedure used in the published analysis \cite{PhysRevD.99.062001}, resulting in 1984 events. Of the selected events, we reserve about 100 events ($\sim$5\% of the data) to perform an independent assessment of the K-shell peak resolution after training. 

    We can increase the size of the training sample by rotating the amplitudes for each event among the channels.
%    Data augmentation is performed to increase the training sample size. 
    We make use of the rotational symmetry of our detector (see Figure~\ref{fig:sfig_channelmap}) by rotating the data of the three inner channels by $0^\circ$, $120^\circ$, or $240^\circ$, and effectively tripling the training data size. The augmented data set thus has rotational symmetry built into it which allows the GBDT to learn this important property of our data.  For SuperCDMS SNOLAB, one can use rotational symmetries combined with symmetry about the vertical axis to flip sides of the detector, thereby increase the training data size by up to a factor of 6 for HV detectors. The trained model is also applied to the other two calibration peaks which served as test data sets.

\begin{figure}[!h]
        \centering
        \begin{subfigure}{0.33\textwidth}
            \centering
            \includegraphics[scale = 0.5]{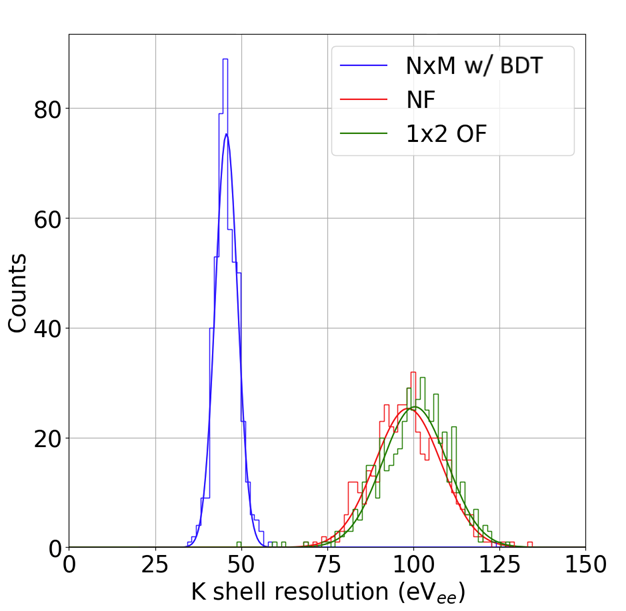}
            \caption{K shell}
            \label{fig:sfig_K}
        \end{subfigure}%
        \begin{subfigure}{0.33\textwidth}
            \centering
            \includegraphics[scale = 0.5]{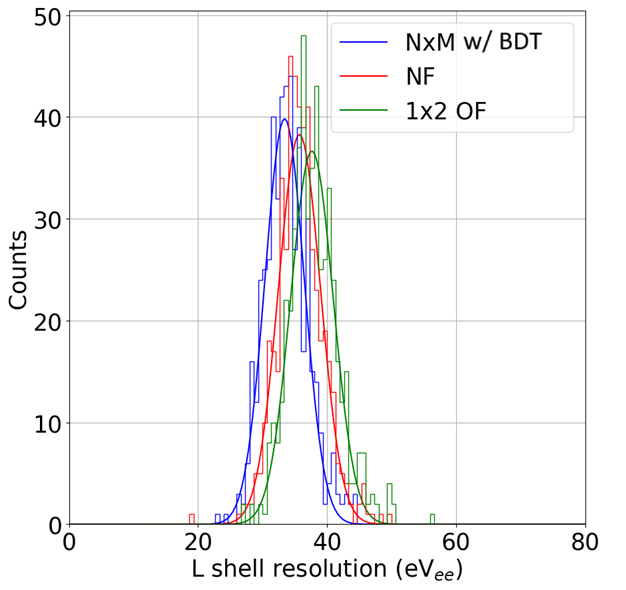}
            \caption{L shell}
            \label{fig:sfig_L}
        \end{subfigure}%
        \begin{subfigure}{0.33\textwidth}
            \centering
            \includegraphics[scale =0.5]{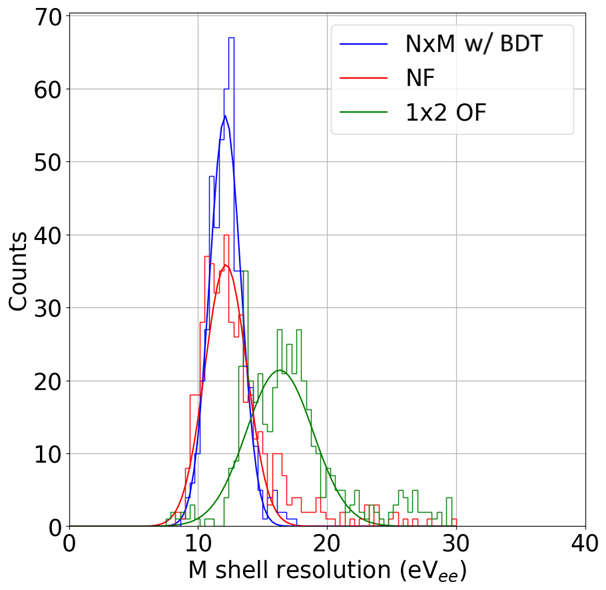}
            \caption{M shell}
            \label{fig:sfig_M}
        \end{subfigure}
        
        \caption{A comparison of the distributions of peak resolutions obtained from 500 bootstrapped samples from various calibration peaks for a Ge detector operated in CDMSlite mode during SuperCDMS Soudan. The $G_\textrm{BDT}$-corrected N$\times$M energy estimates are compared to estimates from other estimators used in SuperCDMS such as the non-stationary filter (NF) and the two-template filter (1$\times$2 OF). }
        \label{fig: CDMSlite calib peaks}
    \end{figure}

    A caveat in this procedure is the limited sample size in the L- and M-shell peaks. To truly assess the efficacy of this technique, the validation must be performed on  independent test samples. 
    Test samples must be independent to validate the machine learning and, separately, to validate the weighted linear-sum estimation of the N$\times$M pipeline.
    %This condition applies both in the machine learning part and the weighted linear sum estimation part of the pipeline used in the N$\times$M. 
    While sufficient training data points ($\sim 5100)$ are available for the K-shell peak, significantly fewer events are available for the L- and M-shell peaks for weight calculation (a factor of ten lower than the K-shell events). Therefore, to produce a fair comparison of the techniques, 
    %we bootstrap the data sets 500 times. 
    we use the ``bootstrap'' method to produce 500 data sets by sampling the event with replacement.    In each bootstrapped data set, 100 events are reserved as an independent test set and the remaining are used to calculate the weights $\alpha_i$ according to Equation~\ref{alpha}. Each bootstrapped sample is then run through the full N$\times$M pipeline and the final calibration peak from the test sample is fit with a Gaussian each time. The resulting distributions of the widths of the calibration peaks are shown in Figure~\ref{fig: CDMSlite calib peaks}. One of the estimators used for comparison is the non-stationary filter, NF, which was explained earlier. The other estimator is the slow component from the 1$\times$2 OF which predominantly contains energy information (described in Section~\ref{sec:template_gen}). Both of these estimators are corrected for linear position dependence using the residual amplitude estimated from the 1$\times$2 filter (see Equation~\ref{E_NF position correction}).  
    
    In Table \ref{CDMSlite resolution table}, we compare the mean energy resolution of the calibration peaks obtained from the fitted Gaussian mean of the bootstrapped distribution of energy resolutions for the three mentioned estimators. We find that the BDT-corrected N$\times$M filter shows the best resolution: about a factor of 2.6 better than the 1$\times$2 estimator for the K-shell peak, up to $\sim$8\% better for the L-shell peak and about 3\% better for the M-shell peak. The small amount of training data available to calculate the weights (a factor ten lower than the K-shell training sample) potentially results in underfitting of the weights for the L and M shells. With larger training samples, resulting in better determination of the weight factors, the resolution performance of the N$\times$M OF for the L- and M-shell peaks could likely be improved. The position uncertainty in CDMS detectors scales as a function of energy and therefore the amount of broadening from position effects decreases with energy. Consequently, the amount of improvement from the BDT correction is expected to decrease with energy.

\begin{table}[h]
	\centering
	% --- Add this line to increase row height ---
	\renewcommand{\arraystretch}{1.5} 
	% -----------------------------------------
	\begin{tabular}{||c|c|c|c||}
		\hline\hline
		\textbf{Estimator} & \textbf{K shell} $\sigma$ \textbf{in eV}$\mathbf{_{ee}}$ & \textbf{L shell} $\sigma$ \textbf{in eV}$\mathbf{_{ee}}$ & \textbf{M shell} $\sigma$ \textbf{in eV}$\mathbf{_{ee}}$ \\
		\hline\hline
		E$^\mathrm{corrected}_{1\times2}$ & 100.42 $\pm$ 0.45 & 37.67 $\pm$ 0.16 & 16.33 $\pm$ 0.14 \\
		\hline
		E$^\mathrm{corrected}_{\rm NF}$  & 98.44 $\pm$ 0.46 & 35.73 $\pm$ 0.15 & 12.17 $\pm$ 0.09\\
		\hline
		E$_{N \times M}$ & 45.73 $\pm$ 0.15 & 33.43 $\pm$ 0.15 & 12.11 $\pm$ 0.06\\
		\hline\hline
	\end{tabular}
	\caption[Comparison of energy resolution of the N$\times$M filter with other SuperCDMS energy estimators along with the 1$\sigma$ statistical uncertainties from the Gaussian fits in Figure~\ref{fig: CDMSlite calib peaks}.]{Comparison of energy resolution of the N$\times$M filter with other SuperCDMS energy estimators in SuperCDMS along with the 1$\sigma$ statistical uncertainties from the Gaussian fits in Figure~\ref{fig: CDMSlite calib peaks}. NF stands for the non-stationary filter. The distribution of the resolutions from 500 bootstrapped samples are used to compare the K-, L- and M-shell peak resolutions.}
	\label{CDMSlite resolution table}
\end{table}

% Scott removes the baseline resolution discussion almost entirely, to see how the paper reads without it
\if(0} 
    Calculation of baseline resolution requires special consideration. Baseline resolution is typically calculated using the OF amplitudes of noise traces, i.e., we ask the question of how well the OF can distinguish a noise event from a real pulse. This calculation requires that we avoid fitting for a time delay, since that biases our OF to the largest random excursions in the noise: OF misinterprets them as pulses. {\color{red} We often use a single template to fit noise traces for the same reason.\footnote{\color{red}Scott: The reasoning for this is {\em still} unclear to me.  I am uncomfortable with this.}} In the case of the N$\times$M filter, this translates to running an N$\times$1 fit using just the lowest order template. We've discussed how the weights, ${\alpha}$, used to estimate $E_\mathrm{lin-est}$ vary as a function of energy and are estimated for each calibration peak, then applied to other energy events via linear interpolation. We cannot use Equation~\ref{alpha} to estimate the zero-energy weights due to $E_\mathrm{true}$ being zero for all events. In the case of our Ge detectors, as is seen in the CDMSlite data set, the M-shell peak is close to our noise threshold. If we make the assumption that the energy-dependent component of the weights is small at the M-shell peak due to its low energy, this allows us to apply the weights estimated from this peak to the baseline within some margin of error. We adopt this approach in this study. {\color{red}We run a 4$\times$4 OF and set the amplitude of components 1, 2 and 3 to zero.\footnote{\color{red}Scott: are you running the fit with those parameters constrained to zero, or are you running it unconstrained and then just replacing the fitted values for the three templates with zero?}} {\color{red} Note that a 4$\times$1 fit will likely do better than a 4$\times$4 fit since all template components are considered in the calculation of the latter, but we adopt the latter approach here to study the effect of multiple templates.\footnote{\color{red}Scott: I have no idea what this sentence is trying to say.}}
    
    The baseline resolution of the detector varied over the course of the CDMSlite run due to changes in environmental noise \cite{PhysRevD.97.022002}. The run was therefore divided into two based on these differences: Run 3a and Run 3b. Table \ref{CDMSlite baseline resolution table} shows the baseline resolution calculated for each sub-run. The average baseline resolution across the full 5 months of data calculated from the table is about 14.26 $\pm$ {\color{red}2.6\footnote{\color{red}Scott: missing a significant digit here.}} eV$_{ee}$ which is statistically consistent with the baseline resolution of 11.86 $\pm$ 2.14 eV$_{ee}$ from a simple 1$\times$1 OF. The unexpectedly higher mean baseline resolution stems from the limitations of our assumption that the linear weights calculated with M-shell peak  can be applied to events with zero energy. More work is required to determine a less error-prone mechanism of evaluating the zero energy weights.

    \begin{table}[h]
        \centering
        \begin{tabular}{||c|c|c||}
            \hline\hline
            \textbf{Estimator} & $\mathbf{\sigma_\mathrm{bs}}$ \textbf{from Run 3a} & $\mathbf{\sigma_\mathrm{bs}}$ \textbf{from Run 3b}  \\
            \hline\hline
            N$\times$M & 12.27 $\pm$ 2.42 eV$_{ee}$ & 16.24 $\pm$ 1.01 eV$_{ee}$ \\
            \hline
            1$\times$1 & 10.62 $\pm$ 2.02 eV$_{ee}$ & 13.1 $\pm$ 0.71 eV$_{ee}$ \\
            \hline\hline
        \end{tabular}
        \caption{Baseline resolution for N$\times$M and 1$\times$1 OFs for each sub-run in CDMSlite. The expected difference between the two baseline resolutions is from better treatment of correlated noise by the N$\times$M filter. The observation is contrary to our expectations in the table due to limitations in estimating $\vec{\alpha}$ for zero energy events.}
        \label{CDMSlite baseline resolution table}
    \end{table}
    
    In SuperCDMS, a resolution model is often constructed as a precursor to performing a sensitivity study or evaluating a dark matter limit. The resolution is affected by three main components:
    \begin{equation}
        \label{Res model}
        \begin{aligned}
            \sigma &= \sqrt{\sigma_\mathrm{bs}^2 + \sigma_F^2 + \sigma_\mathrm{pos}^2}\\
                    &= \sqrt{\sigma_\mathrm{bs}^2 + BE + {(AE)}^2}.
        \end{aligned}
        \label{eq:resmod}
    \end{equation}
    In Equation~\ref{Res model}, $\sigma_\mathrm{bs}$ is the baseline resolution. This is independent of energy. The second term, $\sigma_F$, is called the Fano term which accounts for uncertainties in charge production across the detector and scales as $\sqrt{E}$ with a proportionality constant of $\sqrt{B}$. The final term, $\sigma_\mathrm{pos}$ represents the uncertainty arising from the position of the event. It scales linearly with energy with $A$ being the corresponding proportionality constant. The N$\times$M filter contributes in terms of the first and the last factors. By correcting for correlated noise it reduces $\sigma_\mathrm{bs}$ and the use of GBDT allows us to appropriately correct for position dependence in the detector, thus lowering $\sigma_\mathrm{pos}$. {\color{red}We find that the functional form of Equation~\ref{eq:resmod} is a poor fit for the resolution from the $N\times$M filter, and instead is better fit by the following empirical form:
    \begin{equation}
        \label{Res model NxM}
        \sigma = a\ln{(bE+c)}
    \end{equation}
    where, $a,b$ and $c$ are parameters with units of  eV, eV$^{-1}$ and no units, respectively. This empirical modelling unfortunately has the consequence that it is now non-trivial to interpret this in terms of position uncertainty and a Fano term. However, expanding the logarithm as a Taylor series allows us to determine an energy independent term which is equivalent to the baseline resolution in the conventional model defined in Equation~\ref{eq:resmod}. The Taylor expansion of Equation~\ref{Res model NxM} is:
    \begin{equation}
        \label{Res model Taylor exp}
        \sigma = a\ln(c) + \frac{abE}{c} + ...
    \end{equation}
    The values of $a,b$ and $c$ are estimated with the three calibration peaks and the approximate baseline resolution calculated with the M-shell weights. Squaring Equation~\ref{eq:resmod}, the only constant term is $\sigma_\mathrm{bs}^2$, while the only energy independent term in the square of Equation~\ref{Res model Taylor exp} is $[a\ln(c)]^2$. We interpret this term as the equivalent of the baseline resolution, $\sigma_\mathrm{bs}$. We estimate the parameters, $a,b$ and $c$ separately for each part of the run and the baseline resolution for each part from the empirical formula is shown in Table \ref{CDMSlite empirical model table}. The empirical estimates of the baseline resolution shown in Table \ref{CDMSlite empirical model table} should only be interpreted as theoretical projections for what we could achieve if the weights for the baseline events were calculated perfectly. Thus, we can expect an enhancement in baseline resolution over the 1$\times$1 OF baseline estimate provided we estimate the baseline weights accurately. This is in line with our expectations with the better treatment of correlated noise. More investigation into how these weights ought to be calculated must be performed to produce an in depth understanding of our baseline resolution within the context of the N$\times$M filter.\footnote{\color{red}Scott: this whole section is in my opinion a weak section of the paper.  We can't explain why Eqn 31 fails, and we really don't understand our baseline resolution results.  Is the paper really stronger with this section in it?}}

    \begin{table}[h]
        \centering
        \begin{tabular}{||c|c|c|c|c||}
            \hline\hline
            \textbf{Run} & \textbf{a (eV)} & \textbf{b (eV)} $^{-1}$  & \textbf{c} & $\mathbf{a\ln(c)}$ (eV)\\
            \hline\hline
            3a & 0.0130 & 3.9091 & 1.7686 & 7.4 eV\\
            \hline
            3b & 0.0187 & 1.1723 & 1.8152 & 11.1 eV\\
            \hline\hline
        \end{tabular}
        \caption{Estimates of the parameters in Equation~\ref{Res model NxM} and the energy-independent term in Equation~\ref{Res model Taylor exp} for CDMSlite data.}
        \label{CDMSlite empirical model table}
    \end{table}
\fi % end of the baseline resolutoin and resolution model discussion
    
\subsection{Comment on training samples}
\label{training_samples}

%Most supervised learning algorithms are data hungry: deep learning calls for on the order of millions of data points to produce a usable model \cite{ml-datahungry}. Rare event searches are unfortunately limited in terms of producing large data sets for training. 
Most supervised learning algorithms are data hungry: deep learning typically calls for very large data sets to produce a usable model \cite{ml-datahungry}. Rare event search experiments such as SuperCDMS unfortunately are often limited in the size of the calibration and search data sets available for training.
While we are able to demonstrate significant improvements in energy resolution by our choice of a data-efficient algorithm, we are still limited by small training samples at low energies. Moreover, the training procedure we described in this paper results in data being discarded from the training peak, which is not ideal. Although simulated data sets could in principle be an option to provide independent training data, they are not guaranteed to be fully representative of real-world dark matter data because (a) differences arising from built-in assumptions can skew the simulated training data away from real detector data, and (b) tuning simulation input parameters to match real data sets is an involved process that itself may require large validation data sets. Model training in machine learning requires that the test data set be similar in characteristics and features with the training data for reliable performance. Current research to integrate machine learning into the SuperCDMS data analysis pipeline follows a two-pronged approach: (a) develop more realistic simulated data that closely reproduce detector physics, and (b) explore applicability of machine learning models trained on currently available simulated data sets to existing real data sets.

\section{Conclusion}

We have developed a method for accounting for correlated noise between channels and position-dependent variations in pulse shape in an optimal filter energy estimator.  These effects can be captured by simultaneously fitting data from $N$ channels with $M$ templates derived from principal component analysis, yielding $NM$ amplitudes in total.  A combination of a linear estimator based on these fitted amplitudes and a gradient  boosted decision tree can be used to generate energy or position estimates for events in a SuperCDMS-style detector.  We have demonstrated the validity of this method with simulated data and with data from an older SuperCDMS detector, showing significant improvements in energy resolution.
    
\label{Results and Conclusions}%% The Appendices part is started with the command \appendix;


%% References


\bibliographystyle{elsarticle-num}
\bibliography{paper}






% \end{thebibliography}

\end{document}

%%
%% End of file `ecrc-template.tex'. 